\documentclass[twoside]{article}

\usepackage{blindtext} 

\usepackage[sc]{mathpazo} 
\usepackage[T1]{fontenc} 
\usepackage{microtype} 

\usepackage[english]{babel} 

\usepackage[hmarginratio=1:1,top=32mm,columnsep=20pt]{geometry} 
\usepackage[hang, small,labelfont=bf,up,textfont=it,up]{caption} 
\usepackage{booktabs} 

\usepackage{lettrine} 

\usepackage{enumitem} 
\setlist[itemize]{noitemsep} 

\usepackage{abstract} 

\usepackage{titlesec} 
\renewcommand\thesection{\Roman{section}} 
\renewcommand\thesubsection{\roman{subsection}} 
\titleformat{\section}[block]{\large\scshape\centering}{\thesection.}{1em}{} 
\titleformat{\subsection}[block]{\large}{\thesubsection.}{1em}{} 

\usepackage{titling} 

\usepackage{hyperref} 
\usepackage{authblk}

\usepackage{simplewick}
\usepackage{cite}
\usepackage{slashed}
\usepackage{graphicx}
\usepackage{amsmath}
\usepackage{amssymb}
\usepackage{dsfont}
\usepackage{mathabx}
\usepackage{hyperref}
\usepackage{epigraph}
\usepackage{comment}
\usepackage{bm}
\usepackage[export]{adjustbox}

\newtheorem{defi}{{\textbf{Definitition}}}[section]

\usepackage{enumitem}
\newlist{abbrv}{itemize}{1}
\setlist[abbrv,1]{label=,labelwidth=1in,align=parleft,itemsep=0.1\baselineskip,leftmargin=!}
\usepackage{scalerel,stackengine}
\stackMath
\newcommand\reallywidehat[1]{%
	\savestack{\tmpbox}{\stretchto{%
			\scaleto{%
				\scalerel*[\widthof{\ensuremath{#1}}]{\kern-.6pt\bigwedge\kern-.6pt}%
				{\rule[-\textheight/2]{1ex}{\textheight}}
			}{\textheight}%
		}{0.5ex}}%
	\stackon[1pt]{#1}{\tmpbox}%
}
\stackMath
\newcommand\reallywidecheck[1]{%
	\savestack{\tmpbox}{\stretchto{%
			\scaleto{%
				\scalerel*[\widthof{\ensuremath{#1}}]{\kern-.6pt\bigwedge\kern-.6pt}%
				{\rule[-\textheight/2]{1ex}{\textheight}}
			}{\textheight}%
		}{0.5ex}}%
	\stackon[1pt]{#1}{\scalebox{-1}{\tmpbox}}%
}
\pretitle{\begin{center}\Huge\bfseries} 
\posttitle{\end{center}} 
\title{The causal approach proof for the equivalence of $SDKP_{4}$ and $%
	SQED_{4}$ at tree-level} 
\author[1]{J. Beltran\thanks{c18568@utp.edu.pe, jhosep@ift.unesp.br}}
\author[2]{B. M. Pimentel\thanks{pimentel@ift.unesp.br}}
\author[1]{D. E. Soto\thanks{dsotob@uni.edu.pe}}
\affil[1]{Facultad de Ciencias, Universidad Nacional de Ingeniera (UNI), Avenida Tupac Amaru S/N apartado 31139 Lima, Per\'{u}.}
\affil[2]{S\~{a}o Paulo State University (UNESP), Institute for Theoretical Physics (IFT), R. Dr. Bento Teobaldo Ferraz 271 CEP 01140-070, S\~{a}o Paulo, SP}
\date{\today} 
\renewcommand{\maketitlehookd}{%
\begin{abstract}
\noindent 
The description of the electromagnetic interaction of charged spinless
particles is usually formulated by the Scalar Quantum Electrodynamics.
However, there is an alternative formulation given by the
Duffin-Kemmer-Petiau theory: the Scalar DKP gauge theory. The proof of the
equivalence between these two formulations has been discussed in many researches,
but there is not yet a conclusive proof. In this paper, we initiate a
complete proof in the framework of the Causal Perturbation
theory, showing that both scalar formulations provide the same results for
the differential cross section at tree level.
\end{abstract}
}

\begin{document}

\maketitle


\section{Introduction}
\label{intro}

The Scalar Quantum Electrodynamics (SQED$_4$) is the analogous gauge theory of
the Schwinger-Tomonaga-Feynman Quantum Electrodynamics (QED$_4$) that describes
the electromagnetic interaction of charged spinless particles. As soon as
F.J. Dyson \cite{DysonRad} showed that the QED$_4$ can be consistently
renormalized, in 1950 F. Rohrlich \cite{RohrlichSqed} showed the equivalence
of the Schwinger-Tomonaga theory and Feynman theory for the
SQED case. Furthermore, Rohrlich found the divergence of meson-meson
interaction. Therefore, with the purpose to obtain a renormalized theory, he
assumes that there exist a self-interaction term in the Lagrangian of the
theory.

Moreover, R. Duffin, N. Kemmer and G. Petiau in the period of years between 1936 and 1939
proposed \cite{Petiau, Duffin, Kemmer} a first order equation (known as DKP equation) to describe particles with spin $%
0 $ and spin $1$. Thus, in order to describe the electromagnetic
interaction of charged spinless particles, one can follow the minimal
coupling method for $U\left( 1\right) $ gauge theories. Therefore, it is
possible to construct the interacting theory of DKP's and electromagnetic
fields known as Scalar DKP gauge theory (SDKP$_4$) \cite{Kinoshita1,Kinoshita2,Umezawa,Akh}.

It is known that non-interacting scalar particles can be described by the second order
Klein-Gordon-Fock \cite{KGF} equation and, as we pointed out above, by the
first order DKP equation. Therefore, it is reasonable that in the free
case the DKP and KGF theories are equivalent, both in classical and quantum
pictures. For instance, it was showed that both theories are equivalent in the
classical level for the cases of minimal interaction with electromagnetic 
\cite{ClassElecDKP} and gravitational fields \cite{ClassGravDKP}.

A strict proof of equivalence between both theories, with out self interacting terms, were also given by B.M. Pimentel and V. Ya. Fainberg in \cite{FainPim}. They used the reduction formulas of Lehmann, Symanzik, Zimmermann \cite{redLSZ} to compare the two $S$-matrices constructed for the two theories. In \cite{Fainberg}, they comment that: "In principle, the DKP as well as KGF theories are nonrenormalizable ones even for scalar particles due to the logarithmical
divergence of one loop diagrams of scattering two particles with exchange of
two photons". Then, following the Rohrlich suggestion, to obtain SQED$_4$
renormalizable, they proposed that SDKP$_4$ can becomes renormalizable if we
introduce a self interaction term.

An alternative way to proof the equivalence of SQED$_4$ and SDKP$_4$ is provided by
the Epstein-Glaser perturbative causal approach or Causal Perturbation Theory (CPT) \cite{EG}. The great advantage of this approach is that this is intrinsically finite, therefore,
there is no need to include self-interaction terms neither in SQED$_4$ nor in
SDKP$_4$. For instance, in SQED$_4$, M. D\"{u}tsch, F. Krahe and G. Scharf \cite{Scharf28} found out that there is only need of the derivative coupling
term to reproduce all the well known results of the standard approach.
Furthermore, they demonstrated that SQED$_4$ is (re)normalizable\footnote{In the point of view of CPT there is three kind of theories: Normalizable, non-normalizable and super-normalizable. The prefix re is not used because there are not \textbf{ultraviolet} divergences.} and the self-interacting
term arises naturally from the theory. From these results, we bealieve that CPT is a great framework to analyze
the equivalence of SQED$_4$ and SDKP$_4$, and to determine the (re)normalizability of
the last one.

The origin of CPT started in 1973 when H. Epstein and V. Glaser wrote their article entitled ``The role of locality in perturbation theory"
\cite{EG} where they developed an iterative construction of the $S$-matrix taking as advantage the causal support of the propagators to determine their advanced and retarded part. Ten years later, G. Scharf began to apply the approach to study Quantum Electrodynamics (QED) obtaining a finite theory, in other words, \textbf{a theory without ultraviolet} and infrared divergences \cite{Scharf29,Scharf30,Scharf2,Scharf31A,Scharf31,Scharf32,Scharf33,Scharf34,Scharf5,Scharf1}. From the striking results in QED, G. Scharf and collaborators applied CPT to study other quantum field theories as Yang-Mills \cite{Scharf3,Scharf4,Scharf6,Scharf7,Scharf10,Scharf9}, abelian Higgs mechanism \cite{Scharf8}, Electroweak theory \cite{Scharf12,Scharf13}, Super-symmetry \cite{Scharf14,Scharf15,Scharf17,Scharf21} and Quantum Gravity \cite{Scharf11,Scharf16,Scharf18,Scharf19,Scharf20,Scharf22,Scharf23,Scharf24,Scharf25,Scharf26,Scharf27}. On the other side of the Atlantic ocean, B. M. Pimentel and Collaborators applied CPT in General Quantum Electrodynamics (GQED) \cite{P1A}, Light front Dynamics \cite{P2A,P3A}, SDKP$_4$ \cite{P4A}, gauge Thirring model \cite{P5A,P6A}, and QED$_3$ \cite{P7A,P8A}.

In a previous work \cite{P4A}, Lunardi et al. provides some progress in
the causal approach point of view of SDKP$_4$ oriented to compute the scalar
propagator, the vacuum polarization tensor, the self energy function, and the vertex
correction in the limit of zero momentum transfer. In some sense, this work is a continuation of that one. However, in the present work, we construct the SDKP$_4$ following the \textit{perturbative gauge invariant} property \cite{ScharfG}.

If the SQED$_4$ and SDKP$_4$ approaches are equivalent, both theories must give us the same physical observables. In particular, the scattering cross sections. In this work, we initiate a complete analysis of the SDKP$_4$ in the framework
of CPT. In this sense, we will focus our attention to the Moller and Compton scattering processes to
proof the equivalence at tree level by comparing the differential cross sections.

The paper is organized as follows: In sections \ref{CPT} and \ref{CSP}, CPT will be introduced in generality to be applied to any quantum field theory. In section \ref{QFF}, we develop the quantum properties of free DKP, electromagnetic and fermionic scalar ghost fields to be applied in Scalar Quantum Electrodynamics and to develop gauge invariance at quantum level that will be presented in section \ref{PGI}. The section \ref{S1} is devoted to verified the base term $T_1$, founded in the previous section, by computing the differential cross section for the scattering of scalar particle by a non quantized electromagnetic field. In section \ref{S2}, we will compute the second order causal distribution which contain the Compton and Moller scattering processes and the radiative corrections known as vacuum polarization and self energy function. In sections \ref{S3} and \ref{S4}, we will compute the differential cross sections of the Moller and Compton scattering processes, respectively. Finally in section \ref{DC}, we will present our conclusions.

\section{Causal Perturbation Theory}\label{CPT}

\textbf{The CPT formalism} works directly in the construction of the \emph{scattering operator} as a formal series in the following form
\begin{equation}\label{Definicion de matris s BSEG}
\begin{aligned}
S[g]
&\equiv 1+\sum_{n=1}^{\infty}\frac{1}{n!}\int d^4x_1\ldots d^4x_nT_n(x_1,\ldots,x_n)g(x_1)\ldots g(x_n),\\
\end{aligned}
\end{equation}
where $g(x)\in[0,1]$ is the \emph{switching on-off} Bogoliubov \cite{Bogo1} function and $T_{n}$ are named n-point distributions \cite{Scharf1}. In this approach $g^{\otimes{}n}$ are test functions of $T_{n}$ which are \textit{operator value distributions} (OVD). As we will see later, the distributions $T_{n}$ are well define temporal ordered product of free quantized fields.

In the same sense, we define the inverse of $S$ operator as a formal series too
\begin{equation}\label{iversa de S como una serie}
S^{-1}=\mathds{1}+\sum_{n=1}^\infty\frac{1}{n!}\int d^4x_1\ldots d^4x_n \widetilde{T}_n(x_1,\ldots,x_n)g(x_1)\ldots g(x_n),
\end{equation}
thus, by a formal inversion $S^{-1}=(\mathds{1}+T)^{-1}=1+\sum_{r=1}^\infty(-T)^r$, we obtain the following identity
\begin{equation}\label{T chapeu en funcion de T}
\widetilde{T}_n(x_1,\ldots,x_n)=\sum_{r=1}^{n}(-1)^r\sum_{P_r}T_{n_1}(X_1)\ldots T_{n_r}(X_r),
\end{equation}
where the sum runs over all partitions $P_r$ of the set $X=\{x_1,\ldots,x_n\}$ in $r$ disjoints and not empty sub-sets $X_{i}$.

From (\ref{Definicion de matris s BSEG}), in order to construct the $S$-matrix, we can see that the principal objects to determine are the OVD $T_{n}$. Imposing Poincar\'{e} invariance of $S$, we obtain the following useful properties
\begin{equation}\label{TrasTn} 
T_n(x_1,\ldots,x_n)=T_n(x_1+a,\ldots,x_n+a),
\end{equation}
\begin{equation}\label{Tnlorentzinv}
T(x_1,\ldots,x_n)=T(\Lambda{}x_1,\ldots,\Lambda{}x_n),
\end{equation}
where $\Lambda$ is a Lorentz transformation tensor. \textbf{In CPT, the causality principle is imposed to $S$ from the beginning, so that} we can demonstrate \cite{Scharf1} the following temporal decomposition for $T_{n}$ and $\tilde{T}_{n}$ 
\begin{equation}\label{causalidade de T geral}\begin{aligned}
T_n(x_1,\ldots,x_m,x_{m+1},\ldots,x_n)=T_m(x_1,\ldots,x_m)T_{n-m}(x_{m+1},\ldots,x_n),
\end{aligned}\end{equation}
\begin{equation}\label{DesctiaTn}\begin{aligned}
\widetilde{T}_n(x_1,\ldots,x_m,x_{m+1},\ldots,x_n)=\widetilde{T}_{n-m}(x_{m+1},\ldots,x_n)\widetilde{T}_m(x_1,\ldots,x_m),
\end{aligned}\end{equation}
if and only if $\{x_1^{0},\ldots,x_m^{0}\}>\{x_{m+1}^{0},\ldots,x_n^{0}\}$.

The causal construction of $S$ operator begins defining the first nontrivial term $T_{1}$ with the help of \textit{perturbative gauge invariance} which we will present in section \ref{PGI}.

By considering that we know all distributions $\{T_{n-1},\ldots,T_{1},\tilde{T}_{n-1},\ldots,\tilde{T}_{1}\}$, we can define the \textit{intermediate distributions} $A^{\prime}_{n}$ and $R^{\prime}_{n}$ as
\begin{equation}\label{Aprimagearl}
A'(x_1,\ldots,x_n)\equiv\sum_{P_2}\widetilde{T}_{n_1}(X)T_{n-n_1}(Y,x_n),
\end{equation}
\begin{equation}\label{Rprimagearl}
R'(x_1,\ldots,x_n)\equiv\sum_{P_2}T_{n-n_1}(Y,x_n)\widetilde{T}_{n_1}(X),
\end{equation}
where the sum runs over all partitions $P_2$ of the set $\{x_1,\ldots,x_{n-1}\}$ in two disjoints sub-sets $X$ and $Y$, with $X\neq\emptyset$ as the only restriction. This product is well define because is done between distributions defined in different space points.

The next step is extend the sums (\ref{Aprimagearl}) and (\ref{Rprimagearl}) allowing the empty sub-set $X=\emptyset$
\begin{equation}\label{Agearl}
A_n(x_1,\ldots,x_n)\equiv\sum_{P^0_2}\widetilde{T}_{n_1}(X)T_{n-n_1}(Y,x_n),
\end{equation}
\begin{equation}\label{Rgearl}
R_n(x_1,\ldots,x_n)\equiv\sum_{P^0_2}T_{n-n_1}(Y,x_n)\widetilde{T}_{n_1}(X),
\end{equation}
where $T_{0}=1=\tilde{T}_{0}$ and $P^0_2$ represents the inclusion of the empty set in the partition procedure. It is straightforward to rewrite the sums (\ref{Agearl}) and (\ref{Rgearl}) as
\begin{equation}\label{relacion1}
A_n(x_1,\ldots,x_n)=A'_n(x_1,\ldots,x_n)+T_n(x_1,\ldots,x_n),
\end{equation}
\begin{equation}\label{relacion2}
R_n(x_1,\ldots,x_n)=R'_n(x_1,\ldots,x_n)+T_n(x_1,\ldots,x_n).
\end{equation}

Therefore we can determine the $n$-point distribution by
\begin{equation}\label{SOLTnRES}
T_n(x_1,\ldots,x_n)=
\begin{cases}
A_n(x_1,\ldots,x_n)-A'_n(x_1,\ldots,x_n),\\
R_n(x_1,\ldots,x_n)-R'_n(x_1,\ldots,x_n).
\end{cases}
\end{equation}

In equations (\ref{relacion1}) and (\ref{relacion2}) just $R^{\prime}_{n}$ and $A^{\prime}_{n}$ are known. However, we can demonstrate that $A_n$ and $R_n$ are the advanced and retarded parts of the difference
\begin{equation}\label{DnSpr}\begin{aligned}
D_{n}(x_1,\ldots,x_{n})
\equiv R^{\prime}_{n}(x_1,\ldots,x_{n})-A^{\prime}_{n}(x_1,\ldots,x_{n})\\
=R_{n}(x_1,\ldots,x_{n})-A_{n}(x_1,\ldots,x_{n}),\\
\end{aligned}\end{equation}
where $D_{n}(x_1,\ldots,x_{n})$ is named \textit{causal distribution}. The computation of $A_n$ and $R_n$ from $D_n$ can be done via \textit{ the causal splitting procedure} and it  will be presented in the next section.

\section{Causal splitting procedure}\label{CSP}

In the usual framework, the splitting of a causal distribution in its advanced or retarded part is done by the naively multiplication by the Heaviside step function \cite{Bjorken}. However, this product is not always well defined because in quantum field theory there exist causal singular distributions. As demonstrated by G. Scharf, in QED \cite{Scharf1} this naively procedure were the origin of ultraviolet (UV) divergences.

From the properties of the n-point distributions \textbf{(\ref{T chapeu en funcion de T}-\ref{DesctiaTn})} and by using \textbf{the first} Wick theorem \textbf{\cite{Scharf1}}, it is not difficult to note that the causal distribution $D_n$ is a summation of normal order operator products as follows 
\begin{equation}\label{D2 en funcion de campos libres}
D_n(x_1,\ldots,x_n)=
\sum_k:\prod_j\mathcal{O}(x_j):d^{k}_n(x_1,\ldots,x_n),
\end{equation}
where $\mathcal{O}(x_j)$ represents all kind of OVD and $d^{k}_n(x_1,\ldots,x_n)$ are the numerical part of each summation term obtained via Wick contractions defined as $\contraction{}{\mathcal{O}}{(x_i)}{\mathcal{O}}\mathcal{O}(x_i)\mathcal{O}(x_j)\equiv[\mathcal{O}^{(-)}(x_i),\mathcal{O}^{(+)}(x_j)]$. \textbf{It is useful to represent} (\ref{D2 en funcion de campos libres}) graphically: the non-contracted OVD fields \textbf{resemble} in and out particles, and the numerical distributions $d^{k}_n(x_1,\ldots,x_n)$ represent their connections.

\textbf{The causal split is applied to the numerical distribution} $d_n$. By using the Poincar\'{e} invariance, we can translate $d_n$ by $x_n$
\begin{equation}\label{dtras}
d^{k}_n(x_1,\ldots,x_n)=d^{k}_n(x_1-x_n,\ldots,x_{n-1}-x_n,0)\equiv{}d(\tilde{x}),
\end{equation}
where we define $d(\tilde{x})$ as the general term to denote each numerical distribution to be split and $\tilde{x}=(\tilde{x}_1,\ldots,\tilde{x}_{n-1})$, where $\tilde{x}_i=x_i-x_{n}$. 

\textbf{In the usual treatment of Quantum Field Theory \cite{Bjorken}, the UV divergence problem is ascribed to physical phenomena at very short space-time distance. However, this problem arise because in that treatment the split involved products of} (\ref{dtras}) Heaviside step functions $\Theta(x^{0}_j-x^{0}_n)$ where $j=1,\ldots,n-1$, and such products are ill-defined in the limit $\tilde{x}\rightarrow0$.

\textbf{For a proper treatment we must consider the behavior of} $d({\tilde{x}})$ when points $\{x_1,\ldots,x_{n-1}\}$ are near to $x_n$. This behavior is characterized via a mathematical parameter named \textit{order of singularity} $\omega$ \textbf{of the distribution}\footnote{We will omit the super index $k$ from now.} $d_{n}$. To obtain $\omega$ we first need to determine the \textit{quasi-asymptotic} of $d_{n}$ defined in momentum space\footnote{We use the notations $\hat{f}(p)$ and $\check{f}(p)$ for the direct and inverse Fourier transform, respectively, defined as
$$\hat{f}(x)=(2\pi)^{-\frac{m}{2}}\int{}d^{m}xf(x)e^{ipx},$$$$\check{f}(x)=(2\pi)^{-\frac{m}{2}}\int{}d^{m}xf(x)e^{-ipx}.$$}
 as
\begin{defi}\label{definicion en p de d02}
	A distribution $\hat{d}(p)\in C^{\prime\infty}_{0}(\mathds{R}^m)$ has a quasi-asymptotic $\hat{d}_0(p)$ over $p=\infty$, if for a positive function $\rho(\alpha)$ ($\alpha>0$) the limit
	\begin{equation}\label{limite igual a d03}
	\lim_{\alpha\rightarrow0}\langle\rho(\alpha)\hat{d}(\frac{p}{\alpha}),\check{\psi}(p)\rangle=\langle{}\hat{d}_0(p),\check{\psi}(p)\rangle\neq0,
	\end{equation}
exist for every test function $\check{\psi}(p)$.
\end{defi}

From (\ref{limite igual a d03}), we determine the \textit{power counting function} $\rho(\alpha)$. With $\rho(\alpha)$, we proceed to determine $\omega$ using
\begin{equation}\label{0rho23}
\lim\limits_{\alpha\rightarrow0^{+}}\dfrac{\rho(a\alpha)}{\rho(\alpha)}=a^{\omega}.
\end{equation}

We classify the numerical distributions $d_{n}$ as regular if $\omega<0$, and singular if $\omega\geq0$. When $d_{n}$ is regular, the splitting by using Heaviside step functions are well defined, but in the singular case it is not. In the reference \cite{Scharf1} it is possible to see the demonstration that the numerical retarded part of $d_{2}$ is given by the following formula
\begin{equation}\label{rn23}\begin{aligned}
\hat{r}_{0}(p)
&=\frac{i}{2\pi}Sgn(p^{0})\int\limits_{-\infty}^{\infty}{d}t\frac{\hat{d}(tp)}{1-t+iSgn(p^{0})0^{+}},\quad{}\omega<0;\\
\end{aligned}\end{equation}
\begin{equation}\label{retfmen12No11}\begin{aligned}
\hat{{r}}_{0}(p)
&=\frac{i}{2\pi}Sgn(p^{0})\int{d}t\frac{\hat{d}(tp)}{(t-i0^{+})^{\omega+1}(1-t+iSgn(p^{0})0^{+})}+\sum\limits^{\omega}_{l=0}\hat{C}_{l}p^{l},\quad\omega\geq0;\\
\end{aligned}\end{equation}
where $C_l$ are constants that the \textit{causal splitting procedure} do not fix. They need to be fixed with other physical properties as second order gauge invariance, charge conjugation invariance, etc.

With $\hat{r}_{0}(p)$, we obtain $R_{n}$ and $T_{n}$ by making the necessary substitutions. \textbf{(En mi opinión la parte retardada también debe escribirse explícitamente, similar a (\ref{D2 en funcion de campos libres}))}

\section{Quantized free Fields}\label{QFF}

The free fields are solution of the Lorentz covariant homogeneous field equations at quantum level. In the case of electromagnetic interaction, the value of coupling constant is small enough to expand the  $S$-matrix in terms of free fields.

In this section we develop the necessary properties for electromagnetic an DKP quantized free fields.
\subsection{Electromagnetic field}

The quantized Electromagnetic field $A^{\mu}(x)$ obey the relativistic wave equation \cite{ScharfG}
\begin{equation}\label{EM2WE3}
\Box{}A^{\mu}=0,\quad\Box=g^{\mu\nu}\partial_\nu\partial_\mu,\quad{}g^{\mu\nu}=diag(+,-,-,-).
\end{equation}

Taking into account (\ref{EM2WE3}) as four massless Klein-Gordon-Fock equations, we define the solutions as
\begin{equation}\label{cuantizacion de campo electro componente 02}
A^0(x)=(2\pi)^{-3/2}\int\dfrac{{d}^{3}k}{\sqrt{2\omega}}\left({c}^0(\mathbf{k})e^{-ikx}-{c}^{0}(\mathbf{k})^{\dag}e^{ikx}\right),
\end{equation}
\begin{equation}\label{cuantizacion de campo electro componentes vectoriales2}
A^i(x)=(2\pi)^{-3/2}\int\dfrac{{d}^{3}k}{\sqrt{2\omega}}\left({c}^i(\mathbf{k})e^{-ikx}+{c}^{i}(\mathbf{k})^{\dag}e^{ikx}\right),
\end{equation}
where the operators ${c}^{\mu}(k)^{\dag}$ and ${c}^{\mu}(k)$ are the creation and annihilation operators, respectively, which obey the following commutation relations
\begin{equation}\label{CR2EMFCA}
[{c}^{\mu}(\mathbf{k}),{c}^{\nu}(\mathbf{k}^{\prime})^{\dag}]=
\begin{cases}
\begin{aligned}
\delta(\mathbf{k}-\mathbf{k}^{\prime})\quad \text{for}\quad \mu=\nu,\\
0\quad \text{for}\quad \mu\neq\nu.
\end{aligned}
\end{cases}
\end{equation} 

The minus sign in (\ref{cuantizacion de campo electro componente 02}) has been chosen to have a coherent result in the commutation of two electromagnetic 4-potentials
\footnote{If we do not use the minus sign in (\ref{cuantizacion de campo electro componente 02}), then we would obtain $\left[A^\alpha(x),A^\beta(y)\right]=\delta^\alpha_\beta{i}D_0(x-y)$, which is not correct because we have a second rank Lorentz tensor in the left hand side of the equation and a scalar in the other side.}. This commutation rule is
\begin{equation}\label{conmutador del acmpo electromagnetico forma covariante}
\left[A^\alpha(x),A^\beta(y)\right]=g^{\alpha\beta}iD_0(x-y),
\end{equation}
where $D_{0}(x-y)$ is the massless Lorentz invariant Jordan-Pauli distribution that for $m\neq0$ has the following form
\begin{equation}\label{distribucion jordan pauli final}\begin{aligned}
D_m(x)
&\equiv\frac{i}{(2\pi)^3}\int d^4p \delta({p^2}-m^2)sgn(p^0)e^{-ipx}.\\
\end{aligned}\end{equation}

We define the positive and negative frequency solution for $A^{\mu}$ as
\begin{equation}\label{+A}
A^{\mu(+)}=(2\pi)^{-3/2}\int\dfrac{{d}^{3}k}{\sqrt{2\omega}}{c}^{\mu}(\mathbf{k})^{\dag}e^{ikx}\times
\begin{cases}
\quad1,\quad\text{for $\mu=1,2,3$}\\
-1,\quad\text{for $\mu=0$},
\end{cases}
\end{equation}
\begin{equation}\label{-A}
A^{\mu(-)}=(2\pi)^{-3/2}\int\dfrac{{d}^{3}k}{\sqrt{2\omega}}{c}^\mu(\mathbf{k})e^{-ikx}.
\end{equation}

From (\ref{+A}), (\ref{-A}) and (\ref{CR2EMFCA}), we could compute the following commutation relations
\begin{equation}\label{CR1}
[A^{\mu(-)}(x),A^{\nu(+)}(y)]=g^{\mu\nu}iD_{0}^{(+)}(x-y),
\end{equation}
\begin{equation}\label{CR2}
[A^{\nu(+)}(x),A^{\mu(-)}(y),]=g^{\mu\nu}iD_{0}^{(-)}(x-y),
\end{equation}
where $D_{m=0}^{(+)}(x)$ and $D_{m=0}^{(-)}(x-y)$ are the massless positive and negative parts of Jordan-Pauli distribution
\begin{equation}\label{distribucion escalar positiva final}\begin{aligned}
D^{(+)}_m(x)\equiv\frac{i}{(2\pi)^3}\int{d^{4}}p\delta(p^{2}-m^{2})\Theta(p^{0})e^{-ipx}=\frac{i}{(2\pi)^3}\int\frac{d^3p}{2p^0}e^{-ipx},
\end{aligned}\end{equation}
\begin{equation}\label{distribucion escalar negativa final1}\begin{aligned}
D^{(-)}_m(x)\equiv\frac{-i}{(2\pi)^3}\int{d^{4}}p\delta(p^{2}-m^{2})\Theta(p^{0})e^{ipx}=\frac{-i}{(2\pi)^3}\int\frac{d^3p}{2p^0}e^{ipx},
\end{aligned}\end{equation}
\begin{equation}\label{D+sum}
D_{m}(x)=D^{(+)}_m(x)+D^{(-)}_m(x).
\end{equation}

\subsection{Duffin-Kemmer-Petiau field}

The Duffin-Kemmer-Petiau (DKP) field is a solutions of the DKP equation \cite{Petiau,Duffin,Kemmer}
\begin{equation}\label{e2qpsi}
(i\beta^\mu\partial_\mu-m)\psi(x)=0,
\end{equation}
where $\beta^\mu$ represent four matrices which obey the following algebra
\begin{equation}\label{algebrabeta}
\beta^\mu\beta^\nu\beta^\rho+\beta^\rho\beta^\nu\beta^\mu=\beta^\mu{g}^{\nu\rho}+\beta^\rho{g}^{\mu\nu}.
\end{equation}

The DKP algebra (\ref{algebrabeta}) has three irreducible representations of order 1, 5 and 10. The representation of order 1 is trivial, the next order 5 represent scalar particles and the order 10 represents spin-1 particles. Therefore, to study SDKP$_4$ we will use the representation of order 5.

The DKP field $\psi(x)$ is given by
\begin{equation}\label{DKP geral solution 3}
\psi(x)=\int\frac{d^3p}{(2\pi)^\frac{3}{2}}a(\mathbf{p})u^-(\mathbf{p})e^{-ipx}+
\int\frac{d^3p}{(2\pi)^\frac{3}{2}}b^{\dag}(\mathbf{p})u^+(\mathbf{p})e^{ipx},
\end{equation}
\begin{equation}\label{DKP geral solution 3 conj}\begin{aligned}
\overline{\psi}(x)
&=\int\frac{d^3p}{(2\pi)^\frac{3}{2}}a^{\dag}(\mathbf{p})\overline{u^{-}}(\mathbf{p})e^{ipx}+
\int\frac{d^3p}{(2\pi)^\frac{3}{2}}b(\mathbf{p})\overline{u^{+}}(\mathbf{p})e^{-ipx},\\
\end{aligned}\end{equation}
where the conjugate DKP field $\bar{\psi}(x)$ is obtained via
\begin{equation}\label{conjugadapsi}
\bar{\psi}(x)=\psi^\dag(x)\eta^0, \quad \eta^0=2(\beta^0)^2-1,
\end{equation}
and the operators $a(\mathbf{p})$ and $b(\mathbf{p})$ are the annihilation operators of a scalar particle and antiparticle respectively. They obey the following commutation rules
\begin{equation}\label{comopcreani}\begin{cases}
[a(\mathbf{p}),a^\dag(\mathbf{p}')]=\delta(\mathbf{p}-\mathbf{p}'),\\
[b(\mathbf{p}),b^\dag(\mathbf{p}')]=\delta(\mathbf{p}-\mathbf{p}'),\\
\end{cases}\end{equation}
and null for other ones.

The factors $u^-(\mathbf{p})$ and $u^+(\mathbf{p})$ are column vectors of five elements. To get a positive energy system, they need to be normalized in the following form
\begin{equation}\label{normu}
\overline{u^\pm}\beta^0u^\pm=\mp1.
\end{equation}

From (\ref{DKP geral solution 3}), we can define the positive and negative frequency solutions $\psi^{(+)}$ and $\psi^{(-)}$ as follows
\begin{equation}\label{solfrepos}
\psi^{(+)}(x)\equiv\int\frac{d^3p}{(2\pi)^\frac{3}{2}}b^{\dag}(\mathbf{p})u^+(\mathbf{p})e^{ipx},
\end{equation}
\begin{equation}\label{solfrene}
\psi^{(-)}(x)\equiv\int\frac{d^3p}{(2\pi)^\frac{3}{2}}a(\mathbf{p})u^-(\mathbf{p})e^{-ipx},
\end{equation}
and by conjugation
\begin{equation}\label{solfreposconj}
\overline{\psi}^{(+)}(x)\equiv\int\frac{d^3p}{(2\pi)^\frac{3}{2}}a^{\dag}(\mathbf{p})\overline{u^{-}}(\mathbf{p})e^{ipx},
\end{equation}
\begin{equation}\label{solfreneconj}
\overline{\psi}^{(-)}(x)\equiv\int\frac{d^3p}{(2\pi)^\frac{3}{2}}b(\mathbf{p})\overline{u^{+}}(\mathbf{p})e^{-ipx}.
\end{equation}

For a global $U(1)$ transformation $\delta\psi(x)=ie\alpha\psi(x)$, the conservative Noether current $j^\mu$ is
\begin{equation}\label{4corriente}
j^\mu(x)={e}:\overline{\psi}(x)\beta^{\mu}\psi(x):,
\end{equation}
where $e$ is the unit charge of a scalar particle and the double dots $:\ldots:$ means a normal order product. As usual, the latter is necessary to normalize the vacuum expectation value of the current as $\langle0|j^\mu(x)|0\rangle=0$.

Now, by using (\ref{comopcreani})-(\ref{solfreneconj}), we can determine the following commutations rules
\begin{equation}\label{psi+-subs2dercomS}\begin{aligned}
&[\psi_a^{(-)}(x),\overline{\psi}_b^{(+)}(y)]=\frac{1}{i}S_{ab}^{(+)}(x-y),\\
\end{aligned}\end{equation}
\begin{equation}\label{+-com itreor2}\begin{aligned}
&[\psi_a^{(+)}(x),\overline{\psi}_b^{(-)}(y)]=\frac{1}{i}S^{(-)}_{ab}(x-y),\\
\end{aligned}\end{equation}
where
\begin{equation}\label{S+}
S_{ab}^{(+)}(x)\equiv\frac{1}{m}[i\slashed{\partial}(i\slashed{\partial}+m)]_{ab}D^{(+)}_m(x),
\end{equation}
\begin{equation}\label{S-}\begin{aligned}
S^{(-)}_{ab}(x)&\equiv\frac{1}{m}[i\slashed{\partial}(i\slashed{\partial}+m)]_{ab}D^{(-)}_m(x),\\
\end{aligned}\end{equation}
\begin{equation}\label{JPS}
S(x)\equiv{S}^{(+)}(x)+S^{(-)}(x)=\frac{1}{m}[i\slashed{\partial}(i\slashed{\partial}+m)]D_m(x),\\
\end{equation}
and where the Feynman like notation $\beta^{\mu}\partial_{\mu}=\slashed{\partial}$ is used.

For future applications, we will write here the Fourier transform of the Jordan-Pauli and $S(x)$ distributions. From (\ref{distribucion jordan pauli final}), (\ref{distribucion escalar positiva final}) and (\ref{distribucion escalar negativa final1}), we have
\begin{equation}\label{FD}
\hat{D}_{m}(p)=\frac{i}{2\pi}\delta(p^{2}-m^{2})Sgn(p^{0}),
\end{equation}
\begin{equation}\label{FD+}
\hat{D}_{m}^{(+)}(p)=\frac{i}{2\pi}\delta(p^{2}-m^{2})\Theta(p^{0}),
\end{equation}
\begin{equation}\label{FD-}
\hat{D}_{m}^{(-)}(p)=-\frac{i}{2\pi}\delta(p^{2}-m^{2})\Theta(-p^{0}).
\end{equation}

And from (\ref{S+}), (\ref{S-}) and (\ref{JPS}), it is straightforward to determine the following results
\begin{equation}\label{S+-FT}\begin{aligned}
\hat{S}^{\pm}(p)
&=\frac{1}{m}[\slashed{p}(\slashed{p}+m)]\frac{\pm{}i}{(2\pi)}\Theta(\pm{}p^{0})\delta(p^2-m^2)=\frac{1}{m}[\slashed{p}(\slashed{p}+m)]D_{m}^{\pm}(p),\\
\end{aligned}\end{equation}
\begin{equation}\label{SFT}\begin{aligned}
\hat{S}(p)
&=\frac{1}{m}[\slashed{p}(\slashed{p}+m)]\frac{i}{(2\pi)}Sgn(p^{0})\delta(p^2-m^2)=\frac{1}{m}[\slashed{p}(\slashed{p}+m)]D_{m}(p).\\
\end{aligned}\end{equation}

\section{Perturbative gauge invariance}\label{PGI}

As mention in section \ref{CPT}, to begin the construction of $S$-matrix, we need to define the first nontrivial OVD term $T_{1}(x)$ of (\ref{Definicion de matris s BSEG}). In the regular theory we have $T_1(x)=i:\mathcal{L}_{\text{int}}:$, where $\mathcal{L}_{\text{int}}$ is the interaction Lagrangian. In CPT this is not always true.

As an example, we mention the case of SQED$_4$ constructed with a complex scalar field $\varphi(x)$ which obey the Klein-Gordon-Fock equation \cite{Scharf28}. To obtain a gauge invariant theory, we substitute the partial derivative in the free Lagrangian of $\varphi(x)$ with the covariant derivative $D^{\mu}=\partial^{\mu}+ieA^{\mu}$, obtaining for $\mathcal{L}_{\text{int}}$
\begin{equation}\label{IL}
\mathcal{L}_{\text{int}}=-ieA^{\mu}(\varphi{*}\overleftrightarrow{\partial_{\mu}}\varphi)+e^{2}\varphi^{*}\varphi{}A^{\mu}A_{\mu},
\end{equation}
where $e$ represent the electric charge of scalar particle.

The problem with using the $\mathcal{L}_{\text{int}}$ of (\ref{IL}), to construct $T_1$, is in the second order term $e^{2}\varphi^{*}\varphi{}A^{\mu}A_{\mu}$ which by construction must belong to $T_2$ because in this approach $e$ represents the physical charge and not a mathematical parameter. What is unquestionable is that $T_1$ must be defined from the gauge invariance property but in the quantum level.

As constructed by Scharf et al. in \cite{ScharfG}, a quantum gauge transformation of electromagnetic field has the following form
\begin{equation}\label{QTPG1}
A^{\prime\mu}(x)=A^{\mu}(x)+\lambda{}\partial^{\mu}u(x)+O(\lambda^{2}),
\end{equation}
where $u(x)$ is a quantum free field which obey the massless Klein-Gordon-Fock equation $\Box{}u(x)=0$ and has the following solution
\begin{equation}\label{u}
u(x)\equiv(2\pi)^{-3/2}\int\dfrac{{d}^{3}p}{\sqrt{2\omega}}\left({d}_{2}(\mathbf{p})e^{-ipx}+{d}_{1}(\mathbf{p})^{\dag}e^{ipx}\right).
\end{equation}

By defining the function $\tilde{u}(x)$ as
\begin{equation}\label{tildeu}
\tilde{u}(x)\equiv(2\pi)^{-3/2}\int\dfrac{{d}^{3}p}{\sqrt{2\omega}}\left(-{d}_{1}(\mathbf{p})e^{-ipx}+{d}_{2}(\mathbf{p})^{\dag}e^{ipx}\right),
\end{equation}
and using $\{d_j(\mathbf{p},d_k^{\dag}(\mathbf{q})\}=\delta_{jk}\delta(\mathbf{p}-\mathbf{q})$, we could see that $u(x)$ and $\tilde{u}(x)$ obey the anti-commutation rule
\begin{equation}\label{acr}
\{u(x),\tilde{u}(y)\}=-iD_{0}(x-y).
\end{equation}

The operators $u(x)$ and $\tilde{u}(x)$ are called \textit{scalar fermionic ghost fields} because they obey the massless Klein-Grodon-Fock equation but obey the anti-commutation rule.

The gauge transformation (\ref{QTPG1}) could be obtained in the following form
\begin{equation}\label{QTPG}
A^{\prime\mu}(x)=e^{-i\lambda{}Q}A^{\mu}(x)e^{-i\lambda{}Q},
\end{equation}
where the operator $Q$ is called \textbf{\textit{gauge charge}}. Expanding the exponential, we obtain
\begin{equation}\label{QTPG2}
A^{\prime\mu}(x)=A^{\mu}(x)-i\lambda[Q,A^{\mu}(x)]+O(\lambda^{2}),
\end{equation}
then, comparing the equations (\ref{QTPG1}) and (\ref{QTPG2}), we obtain a differential operator equation for $Q$
\begin{equation}\label{EQ}
[Q,A^{\mu}(x)]=i\partial^{\mu}u(x).
\end{equation}

The solution for $Q$ in (\ref{EQ}) is
\begin{equation}\label{QIS}
Q=\int{d}^{3}x[\partial_{\nu}A^{\nu}\partial_0u-(\partial_0\partial_{\nu}A^{\nu})u]=\int{d}^{3}x\partial_{\nu}A^{\nu}\overleftrightarrow{\partial}_{0}u,
\end{equation}
where the integral is performed over a hyperplane $x^{0}=\text{constant}$.

Now, we define the \textit{gauge derivative} $d_{Q}$ as
\begin{equation}\label{DQ}
d_{Q}F\equiv[Q,F],\quad{}d_{Q}G\equiv\{Q,G\},
\end{equation}
where $F$ is a product of Bose fields by an even number of fermionic ghost fields, and $G$ is a product of a Bose fields by an odd number of fermionic ghost fields, 

In order to obtain a gauge invariant theory, we demand that all n-point distributions $T_{n}$ obey the following identity
\begin{equation}\label{TnP}
d_{Q}T_{n}(x_{1},\ldots,x_{n})=i\sum\limits_{l=1}^{n}\frac{\partial}{\partial{}x_{l}^{\mu}}T^{\mu}_{n/l}(x_{1},\ldots,x_{n}),
\end{equation}
where $T^{\mu}_{n/l}(x_{1},\ldots,x_{n})$ is the following well defined time ordering product constructed via CPT
\begin{equation}\label{Tnl}
T^{\mu}_{n/l}(x_{1},\ldots,x_{n})=T\{T_{1}(x_{1})\ldots{}T_{1/1}^{\mu}(x_{l})\ldots{}T_{1}(x_{n})\},
\end{equation}
and $T_{1/1}^{\mu}$ is called the Q-vertex. The identity (\ref{TnP}) is called \textbf{\textit{perturbative gauge invariance condition}}.

For one massless gauge field $A^{\mu}(x)$ we have $Q$ in the form (\ref{QIS}) and the following gauge transformations
\begin{equation}\label{GT1}
d_{Q}A^{\mu}(x)=i\partial^{\mu}u(x),
\end{equation}
\begin{equation}\label{GT2}
d_{Q}u(x)=0,
\end{equation}
\begin{equation}\label{GT3}
d_{Q}\tilde{u}(x)=-i\partial_{\mu}A^{\mu}(x).
\end{equation}

First of all, to determine $T_{1}(x)$ we can use (\ref{TnP}) for $n=1$
\begin{equation}\label{T1G}
d_{Q}T_{1}(x_{1})=i\partial_{\mu}T^{\mu}_{1/1}(x_{1}).
\end{equation}

Secondly, because of the adiabatic limit $g(x)\rightarrow{}1$, the term $T_{1}(x)$ must contain all kind of interactions between gauge and matter fields. As show by Scharf et al. \cite{Scharf3,Scharf4,Scharf6,Scharf7}, only in the case where there are a collection of massless gauge fields $A^{\mu}_{a}(x)$, where $a=1,\ldots,N$, the condition (\ref{T1G}) allows the existence of self interaction terms between gauge and ghost fields. 

For SDKP, where we have one massless gauge field, $T_{1}$ only contain the  interaction between electromagnetic and matter current in the form
\begin{equation}\label{T12}
T_{1}^{(\text{SDKP})}(x_1)=ij^{\mu}(x_1)A_{\mu}(x_1).
\end{equation}

Because $j^{\mu}(x)$ represents the matter current, it contains DKP fields $\psi(x)$ and $\bar{\psi}(x)$. Therefore, with the help of (\ref{GT1}), (\ref{GT2}) and ((\ref{GT3})), by taking the gauge derivative of (\ref{T12}), we have
\begin{equation}\label{GDTDKP}
d_{Q}T_{1}^{(\text{SDKP})}(x_{1})=d_{Q}\{ij^{\mu}(x_1)A_{\mu}(x_1)\}=ij^{\mu}(x_1)d_{Q}A_{\mu}(x_1)=-j^{\mu}(x_1)\partial_{\mu}u(x_1).
\end{equation}

From (\ref{T12}) and (\ref{GDTDKP}), we can conclude that to satisfy the condition (\ref{T1G}), the matter current $j^{\mu}(x)$ must be the divergenceless Noether current (\ref{4corriente}), therefore
\begin{equation}\label{CC}
\partial_{\mu}j^{\mu}(x)=0\Longrightarrow{}d_{Q}T_{1}^{(\text{SDKP})}(x_{1})=i\partial_{\mu}T_{1/1}^{\mu(\text{SDKP})}=i\partial_\mu(ij^{\mu}(x_1)u(x_1)).
\end{equation}

Finally, replacing (\ref{4corriente}) in the expressions for the $Q$-vertex, we obtain for it and $T_{1}$ the following forms
\begin{equation}\label{SDKPT11}
T_{1/1}^{\mu(\text{SDKP})}=ij^{\mu}(x_1)u(x_1)=i{e}:\overline{\psi}(x_1)\beta^{\mu}\psi(x_1):u(x_1),
\end{equation}
\begin{equation}\label{T13}
T_{1}^{(\text{SDKP})}(x)=i{e}:\overline{\psi}(x_1)\beta^{\mu}\psi(x_1):A_{\mu}(x_1).
\end{equation}

\section{Scattering of DKP particle by static external field}\label{S1}

As a first application, we are going to determine the differential cross section $d\sigma/d\Omega$ in the scattering of scalar by an external non-quantized electromagnetic field $A_\mu^{ext}$.

If the system includes $A_\mu^{ext}$, then we need to perform the substitution $A_{\mu}\rightarrow A_\mu+A_\mu^{ext}$ in the construction of the $S$-Matrix. Therefore, the perturbative expansion of $S$-matrix includes a term
\begin{equation}\label{Sextterm}
S=\ldots+\int{d}^4x{i}e:\overline{\psi}(x)\beta^\mu\psi(x):A_\mu^{ext}(x)+\ldots.
\end{equation}
This term is the important one to study a scattered scalar particle by $A_\mu^{ext}$. By considering that the initial and final states does not include the creation (or annihilation) of photons, they take the following form
\begin{equation}\label{estado inicial}
\begin{aligned}
|in\rangle&=|\Psi_i\rangle=\int{}d^{3}p_{1}\Phi_{i}(\mathbf{p}_{1})a^\dag(\mathbf{p}_1)|0\rangle,
\end{aligned}
\end{equation}
\begin{equation}\label{estado final}
\begin{aligned}
|out\rangle&=|\Psi_f\rangle=\int{}d^{3}p_{2}\Phi_{f}(\mathbf{p}_{2})a^\dag(\mathbf{p}_2)|0\rangle,
\end{aligned}
\end{equation}
where $\mathrm{a}^\dag(\mathbf{p}_{1,2})$ are the creation operators for scalar particles with momentums $\mathbf{p}_{1,2}$. The functions $\Phi_{i}$ and $\Phi_{f}$ are wave packets sharply peaked in $\mathbf{p}_{i}$ and $\mathbf{p}_{f}$, respectively.

Now, by computing the scattering amplitude $\mathcal{A}_{if}=\langle\Psi_{f}|S|\Psi_{i}\rangle$, it is not difficult to see that the only non-null result is
\begin{equation}\label{Sifex}\begin{aligned}
\mathcal{A}_{if}
&=ie\int{}d^4x\langle\Psi_f|:\overline{\psi}(x)\beta^\mu\psi(x):|\Psi_i\rangle{}A^{ext}_\mu(x)\\
&=ie\int{}d^4x\int{}d^{3}p_{2}\int{}d^{3}p_{1}\Phi^{*}_{f}(\mathbf{p}_{2})\Phi_{i}(\mathbf{p}_{1})\times\\
&\quad\times\langle0|a(\mathbf{p}_2):\overline{\psi}(x)\beta^\mu\psi(x):a^\dag(\mathbf{p}_1)|0\rangle{}A^{ext}_\mu(x).\\
\end{aligned}\end{equation}

In order to compute the term $\langle0|a(\mathbf{p}_2):\overline{\psi}(x)\beta^\mu\psi(x):a^\dag(\mathbf{p}_1)|0\rangle$, we can use the Wick theorem. Therefore, only the term with the two simultaneous contractions $\contraction{}{a}{(\mathbf{p}_2)}{\bar{\psi}(x}a(\mathbf{p}_2)\bar{\psi}(x)$ and $\contraction{}{\psi}{(x)}{a}\psi(x)a^\dag(\mathbf{p}_1)$ contributes to (\ref{Sifex}). These contractions are
\begin{equation}\label{cont1A}\begin{aligned}
\contraction{}{{a}}{(\mathbf{p}_2)}{\overline\psi}
{a}(\mathbf{p}_2)\overline{\psi}(x)
\contraction{=}{{a}}{(\mathbf{p}_2)}{\overline{\psi}}
&={a}(\mathbf{p}_2)\overline{\psi}^{(+)}(x)
\contraction[3ex]{=}{{a}}{(\mathbf{q}_f)\int\frac{d^3p}{(2\pi)^\frac{3}{2}}}{a}
={a}(\mathbf{q}_2)\int\frac{d^3p}{(2\pi)^\frac{3}{2}}a^\dag(\mathbf{p})\overline{u^{-}}(\mathbf{p})e^{ipx}\\
&=\int\frac{d^3p}{(2\pi)^\frac{3}{2}}\delta(\mathbf{p}_2-\mathbf{p})\overline{u^{-}}(\mathbf{p})e^{ipx}
=\frac{1}{(2\pi)^\frac{3}{2}}\overline{u^{-}}(\mathbf{p}_2)e^{ip_2x},\\
\end{aligned}\end{equation}
\begin{equation}\label{cont2A}\begin{aligned}
\contraction{}{\psi}{(x)}{a}
\psi(x)a^\dag(\mathbf{p}_1)
\contraction{=}{\psi}{^{(-)}(x)}{{a}}
&=\psi^{(-)}(x){a}^\dag(\mathbf{p}_1)
\contraction{=\int\frac{d^3p}{(2\pi)^\frac{3}{2}}}{a}{(\mathbf{p})u^-(\mathbf{p})e^{-ipx}}{{a}}
=\int\frac{d^3p}{(2\pi)^\frac{3}{2}}a(\mathbf{p})u^-(\mathbf{p})e^{-ipx}{a}^\dag(\mathbf{p}_1)\\
&=\int\frac{d^3p}{(2\pi)^\frac{3}{2}}\delta(\mathbf{p}_1-\mathbf{p})u^-(\mathbf{p})e^{-ipx}
=\frac{1}{(2\pi)^\frac{3}{2}}u^-(\mathbf{p}_1)e^{-ip_1x}.\\
\end{aligned}\end{equation}

With the help of (\ref{cont1A}) and (\ref{cont2A}), the expression (\ref{Sifex}) takes the following form
\begin{equation}\label{Sifex22}\begin{aligned}
\mathcal{A}_{if}
&=ie\int{}d^{3}p_{2}\int{}d^{3}p_{1}\Phi^{*}_{f}(\mathbf{p}_{2})\Phi_{i}(\mathbf{p}_{1})\frac{1}{(2\pi)^{3}}\overline{u^{-}}(\mathbf{p}_2)\beta^\mu u^-(\mathbf{p}_1)\int{}d^4xe^{-i(p_1-p_2)x}A^{ext}_\mu(x)\\
\end{aligned}\end{equation}

If the electromagnetic field is an static one, then we can replace $A^{ext}_\mu(x)=A^{ext}_\mu(\mathbf{x})$. The latter allows us to evaluate the integral in $x^{0}$ to obtain
\begin{equation}\label{Sifex23}\begin{aligned}
\mathcal{A}_{if}
&=\int{}d^{3}p_{2}\int{}d^{3}p_{1}\Phi^{*}_{f}(\mathbf{p}_{2})\Phi_{i}(\mathbf{p}_{1})\mathcal{M}_{if}(\mathbf{p}_1,\mathbf{p}_2)\delta(E_1-E_2),\\
\end{aligned}\end{equation}
where
\begin{equation}\label{Mext}
\mathcal{M}_{if}(\mathbf{p}_1,\mathbf{p}_2)=\frac{ie}{(2\pi)^{\frac{1}{2}}}\overline{u^{-}}(\mathbf{p}_2)\beta^\mu{}u^-(\mathbf{p}_1)\hat{A}(\mathbf{p}_2-\mathbf{p}_1),
\end{equation}
\begin{equation}\label{Atrans}
\hat{A}(\mathbf{p}_2-\mathbf{p}_1)
=(2\pi)^{-\frac{3}{2}}\int{}d^3xe^{-i(\mathbf{p}_2-\mathbf{p}_1)\mathbf{x}}A_\mu(\mathbf{x}).
\end{equation}

With the computation of $\mathcal{A}_{if}$, we can determine the transition probability $P_{if}$ defined as
\begin{equation}\label{probabilidad de transicionex0}\begin{aligned}
P_{if}
&=|\mathcal{A}_{if}|^2.\\
\end{aligned}\end{equation}

Replacing (\ref{Sifex23}) into (\ref{probabilidad de transicionex0}), we have
\begin{equation}\label{probabilidad de transicionex}\begin{aligned}
P_{if}
&=\int{d}^3p_1{d}^3p_2\Phi_f(\mathbf{p}_2)\tilde{S}^*_{if}(\mathbf{p}_1,\mathbf{p}_2)\Phi^*_i(\mathbf{p}_1)
\int{d}^3p'_1{d}^3p'_2\Phi^*_f(\mathbf{p}'_2)\tilde{S}_{if}(\mathbf{p}'_1,\mathbf{p}'_2)\Phi_i(\mathbf{p}'_1).\\
\end{aligned}\end{equation}
where
\begin{equation}\label{Tildesif}
\tilde{S}_{if}=\mathcal{M}_{if}(\mathbf{p}_1,\mathbf{p}_2)\delta(E_1-E_2)
\end{equation}

By summing over all possible final states, we obtain
\begin{equation}\label{probabilidad de transicionex2}\begin{aligned}
\sum_{f}P_{if}
&=\int{d}^3p_1{d}^3p_2\mathcal{M}^*_{if}(\mathbf{p}_1,\mathbf{p}_2)\delta(E_{1}-E_{2})\Phi^*_i(\mathbf{p}_1)
\int{d}^3p'_1\mathcal{M}_{if}(\mathbf{p}'_1,\mathbf{p}_2)\delta(E'_{1}-E_{2})\Phi_i(\mathbf{p}'_1)\\
\end{aligned}\end{equation}

From the fact that the function $\Phi_i(\mathbf{p}_1)$ is sharply peaked in $p_i$ and by considering that the width of $\Phi_i(\mathbf{p}_1)$ is too small compared with the scale of varying of $\mathcal{M}_{if}$, we could rewrite (\ref{probabilidad de transicionex2}) as follows
\begin{equation}\label{probabilidad de transicionex3}\begin{aligned}
\sum_{f}P_{if}
&=\int{d}^3p_2|\mathcal{M}_{if}(\mathbf{p}_i,\mathbf{p}_2)|^2
\int{d}^3p_1{d}^3p'_1\delta(E_{1}-E_{2})\Phi^*_i(\mathbf{p}_1)\delta(E'_{1}-E_{2})\Phi_i(\mathbf{p}'_1).\\
\end{aligned}\end{equation}

In order to reduce (\ref{probabilidad de transicionex3}), we must evaluate the $\mathbf{p}_2$ integral in spherical coordinates and by replacing the integral form of the delta function $\delta(E'_{1}-E_{1})=(2\pi)^{-1}\int{dt}e^{-i(E'_{1}-E_{1})t}$, we obtain
\begin{equation}\label{probabilidad de transicionex4B}\begin{aligned}
\sum_{f}P_{if}
&=\int{d}\Omega_2|\mathcal{M}_{if}(\mathbf{p}_i,\mathbf{p}_2)|^2|\mathbf{p}_i|E_{i}(2\pi)^{2}\int{dt}|\Phi(t,\mathbf{x}=\mathbf{0})|^2,\\
\end{aligned}\end{equation}
where $\Phi(t,\mathbf{x})$ is the following free wave packet function in configuration space
\begin{equation}\label{wavex}
\Phi(t,\mathbf{x})=(2\pi)^{-\frac{3}{2}}\int{d}^3q\Phi_i(\mathbf{q})e^{-i(E_{q}t-\mathbf{q}\mathbf{x})}.
\end{equation}

After considering that the velocity of the scattered incoming particles is $\mathbf{v}$, the wave packet takes the following form
\begin{equation}\label{shifted}
\Phi(t,\mathbf{x})=\Phi_0(\mathbf{x}+\mathbf{v}t).
\end{equation}

Now, by averaging (\ref{probabilidad de transicionex4B}) in a cylinder of radius $R$ parallel to $\mathbf{v}$ and regarding the wave packet (\ref{shifted}), we have
\begin{equation}\label{probabilidad de transicionex5}\begin{aligned}
\sum_{f}P_{if}(R)
&=\frac{1}{\pi{}R^2}\int\limits_{x_{\perp}\leq{}R}{d}^2x_{\perp}\int{dt}|\Phi_0(\mathbf{x}+\mathbf{v}t)|^2\int{d}\Omega_2|\mathcal{M}_{if}(\mathbf{p}_i,\mathbf{p}_2)|^2|\mathbf{p}_i|E_{i}(2\pi)^{2}.\\
\end{aligned}\end{equation}

As usual, the cross section is defined as
\begin{equation}\label{crossex}
\sigma=\lim\limits_{R\rightarrow\infty}\pi{}R^2\sum_{f}P_{if}(R).
\end{equation}

Because of the limit $R\rightarrow\infty$, the first two integrals in (\ref{probabilidad de transicionex5}) can be performed as follows
\begin{equation}\label{Int}
\int\limits_{x_{\perp}\leq{}R}{d}^2x_{\perp}\int{dt}|\Phi_0(\mathbf{x}+\mathbf{v}t)|^2=\frac{1}{|\mathbf{v}|}\int{}dx^3|\Phi_0|^2=\frac{1}{|\mathbf{v}|}
\end{equation}

Therefore, after replacing (\ref{probabilidad de transicionex5}) into (\ref{crossex}), we get
\begin{equation}\label{crossex2A}\begin{aligned}
\sigma
&=E^2_{p_i}(2\pi)^{2}\int{d}\Omega_2|\mathcal{M}_{if}(\mathbf{p}_i,\mathbf{p}_2)|^2.\\
\end{aligned}\end{equation}

From (\ref{crossex2A}), it is straightforward to obtain the differential cross section as
\begin{equation}\label{exdiffcrosssec}
\frac{d\sigma}{d\Omega}=(2\pi)^2E^2_i|\mathcal{M}(\mathbf{p}_f,\mathbf{p}_i)|^2.
\end{equation}

Replacing (\ref{Mext}) into (\ref{exdiffcrosssec}), we get
\begin{equation}\label{exdiffcrosssec2}\begin{aligned}
\frac{d\sigma}{d\Omega}
&=\frac{(2\pi)E^2_ie^2}{4m^2{p_i}^0{p_f}^0}
Tr[\slashed{p}_{f}(\slashed{p}_{f}+m)\beta^\mu\slashed{p}_{i}(\slashed{p}_{i}+m)\beta^\nu]
\hat{A}_{\nu}(\mathbf{p}_f-\mathbf{p}_i)\hat{A}_{\mu}(\mathbf{p}_i-\mathbf{p}_f)\\
&=\frac{(2\pi)E^2_ie^2}{4{p_i}^0{p_f}^0}
[{p}_{i}^\mu{p}_{i}^\nu+{p}_{f}^\nu{p}_{f}^\mu+{p}_{f}^\mu{p}_{i}^\nu+{p}_{f}^\nu{p}_{i}^\mu]
\hat{A}_{\nu}(\mathbf{p}_f-\mathbf{p}_i)\hat{A}_{\mu}(\mathbf{p}_i-\mathbf{p}_f)\\
\end{aligned}\end{equation}
where we used the following trace identities
\begin{equation}\label{TRAZ1}
Tr[\beta^{\mu_{1}}\beta^{\mu_{2}}\ldots\beta^{\mu_{2n-1}}]=0,
\end{equation}
\begin{equation}\label{TRAZ2}
Tr[\beta^{\mu_{1}}\beta^{\mu_{2}}\ldots\beta^{\mu_{2n}}]=g^{\mu_{1}\mu_{2}}g^{\mu_{3}\mu_{4}}\ldots{}g^{\mu_{2n-1}\mu_{2n}}+g^{\mu_{2}\mu_{3}}g^{\mu_{4}\mu_{5}}\ldots{}g^{\mu_{2n}\mu_{1}}.
\end{equation}

Regarding the coulomb potential where only the $0$-component is not zero
\begin{equation}\label{Coulomb Potential}
A^{0}(\mathbf{x})=\frac{Ze}{|\mathbf{x}|},\quad \hat{A}^{0}(\mathbf{p})=\sqrt{\frac{2}{\pi}}\frac{Ze}{|\mathbf{p}|^2},
\end{equation}
we have for (\ref{exdiffcrosssec2}) the following result
\begin{equation}\label{exdiffcrosssec3}\begin{aligned}
\frac{d\sigma}{d\Omega}
&=\frac{(2\pi)E^2_ie^2}{4{p_i}^0{p_f}^0}
[{p}_{i}^0{p}_{i}^0+{p}_{f}^0{p}_{f}^0+{p}_{f}^0{p}_{i}^0+{p}_{f}^0{p}_{i}^0]
\hat{A}_{0}(\mathbf{p}_f-\mathbf{p}_i)\hat{A}_{0}(\mathbf{p}_i-\mathbf{p}_f)\\
&={Z^2e^4}\frac{E^2}{4|\mathbf{p}|^4\sin^4(\vartheta/2)}.\\
\end{aligned}\end{equation}

The latter formula is equivalent to that obtained in \cite{Akh} via Feynman diagrammatically approach.

\section{Causal Distribution in second order $D_2(x,y)$}\label{S2}

After setting $T_{1}(x)$, the second step is to compute the causal distribution $D_2(x,y)$. By following (\ref{Aprimagearl}) and (\ref{Rprimagearl}), the intermediate distributions in second order $A'_2$ and $R'_2$ are given by the following expressions
\begin{equation}\label{Ap}
A'_2(x,y)=\tilde{T}_1(x)T_1(y)=-{T}_1(x)T_1(y),
\end{equation}
\begin{equation}\label{Rp}
R'_2(x,y)=T_1(y)\tilde{T}_1(x)=-T_1(y){T}_1(x),
\end{equation}
where (\ref{T chapeu en funcion de T}) is used.

Replacing (\ref{T13}) into (\ref{Ap}) and (\ref{Rp}), and using Wick theorem to obtain normal ordered terms, we have
\begin{equation}\label{A2primacalc}\begin{aligned}
A'_2(x,y)
&=-T_1(x)T_1(y)\\
&=e^2:\overline{\psi}_a(x)\beta^\mu_{ab}\psi_b(x)::\overline{\psi}_c(y)\beta^\nu_{cd}\psi_d(y):A_\mu(x)A_\nu(y)\\
\contraction{=[e^2:}{\overline{\psi}}{_a(x)\beta^\mu_{ab}\psi_b(x)\overline{\psi}_c(y)\beta^\nu_{cd}}{\psi}
&=e^2\Big[:\overline{\psi}_a(x)\beta^\mu_{ab}\psi_b(x)\overline{\psi}_c(y)\beta^\nu_{cd}\psi_d(y):
\contraction{+:\overline{\psi}_a(x)\beta^\mu_{ab}}{\psi}{_b(x)}{\overline{\psi}}
+:\overline{\psi}_a(x)\beta^\mu_{ab}\psi_b(x)\overline{\psi}_c(y)\beta^\nu_{cd}\psi_d(y):+\\
\contraction{\quad+:}{\overline{\psi}}{_a(x)\beta^\mu_{ab}\psi_b(x)\overline{\psi}_c(y)\beta^\nu_{cd}}{\psi}
\bcontraction{\quad+:\overline{\psi}_a(x)\beta^\mu_{ab}}{\psi}{_b(x)}{\overline{\psi}}
&\quad+:\overline{\psi}_a(x)\beta^\mu_{ab}\psi_b(x)\overline{\psi}_c(y)\beta^\nu_{cd}\psi_d(y):
+:\overline{\psi}_a(x)\beta^\mu_{ab}\psi_b(x)\overline{\psi}_c(y)\beta^\nu_{cd}\psi_d(y):\Big]\times\\
\contraction{\quad\times\big[:A_\mu(x)A_\nu(y):+}{A}{_\mu(x)}{A}
&\quad\times\Big[:A_\mu(x)A_\nu(y):+A_\mu(x)A_\nu(y)\Big],\\
\end{aligned}\end{equation}
where the field contractions are
\begin{equation}\label{CR12}
\contraction{}{A}{^{\mu}(x)}{A}
A^{\mu}(x)A^{\nu}(y)=[A^{\mu(-)}(x),A^{\nu(+)}(y)]=g^{\mu\nu}iD_{0}^{(+)}(x-y),
\end{equation}
\begin{equation}\label{CR13}
\contraction{}{{\psi}_{a}}{(x)}{\bar{\psi}}
{\psi}_{a}(x)\bar{\psi}_{b}(y)=[\psi^{(-)}(x),\bar{\psi}^{(+)}(y)]=\frac{1}{i}S_{ab}^{(+)}(x-y),
\end{equation}
\begin{equation}\label{CR14}
\contraction{}{\bar{\psi}}{_{c}(x)}{{\psi}}
\bar{\psi}_{c}(x){\psi}_{d}(y)=[\bar{\psi}^{(-)}(x),{\psi}^{(+)}(y)]=-\frac{1}{i}S^{(-)}_{dc}(y-x).
\end{equation}

Replacing (\ref{CR12}), (\ref{CR13}) and (\ref{CR14}) into (\ref{A2primacalc}), we obtain
\begin{equation}\label{A2primacalc2a}\begin{aligned}
A'_2(x,y)
&=A^{\prime(1)}_2(x,y)+A^{\prime(2)}_2(x,y)+A^{\prime(3)}_2(x,y)+A^{\prime(4)}_2(x,y)+A^{\prime(5)}_2(x,y)+\\
&\quad+e^2:\overline{\psi}\beta^\mu(x)\psi(x)\overline{\psi}(y)\beta^\nu\psi(y)::A_\mu(x)A_\nu(y):,\\
\end{aligned}\end{equation}
where
\begin{equation}\label{aprima1}\begin{aligned}
A^{\prime(1)}_2(x,y)
&=ie^2g_{\mu\nu}:\overline{\psi}(x)\beta^\mu\psi(x)D_0^{(+)}(x-y)\overline{\psi}(y)\beta^\nu\psi(y):,\\
\end{aligned}\end{equation}
\begin{equation}\label{aprima2}\begin{aligned}
A^{\prime(2)}_2(x,y)
&=ie^2:\overline{\psi}(y)\beta^{\nu}S^{(-)}(y-x)\beta^\mu\psi(x)::A_\mu(x)A_\nu(y):-\\
&\quad-ie^2:\overline{\psi}(x)\beta^\mu{}S^{(+)}(x-y)\beta^\nu\psi(y)::A_\mu(x)A_\nu(y):,\\
\end{aligned}\end{equation}
\begin{equation}\label{aprima3}\begin{aligned}
A^{\prime(3)}_2(x,y)
&=e^2:A_\mu(x)Tr[\beta^{\mu}S^{(+)}(x-y)\beta^{\nu}S^{(-)}(y-x)]A_\nu(y):,\\
\end{aligned}\end{equation}
\begin{equation}\label{aprima4}\begin{aligned}
A^{\prime(4)}_2(x,y)
&=-e^2g_{\mu\nu}:\overline{\psi}(y)\beta^{\nu}S^{(-)}(y-x)D_0^{(+)}(x-y)\beta^\mu\psi(x):+\\
&\quad+e^2g_{\mu\nu}:\overline{\psi}(x)\beta^{\mu}S^{(+)}(x-y)D_0^{(+)}(x-y)\beta^{\nu}\psi(y):,\\
\end{aligned}\end{equation}
\begin{equation}\label{aprima5}\begin{aligned}
A^{\prime(5)}_2(x,y)
&=ie^2g_{\mu\nu}D_0^{(+)}(x-y)Tr[\beta^{\mu}S^{(+)}(x-y)\beta^{\nu}S^{(-)}(y-x)].\\
\end{aligned}\end{equation}

Similarly, for $R'_{2}(x,y)$ we have
\begin{equation}\label{R2primacalc2a}\begin{aligned}
R'_2(x,y)
&=R^{\prime(1)}_2(y,x)+R^{\prime(2)}_2(y,x)+R^{\prime(3)}_2(y,x)+R^{\prime(4)}_2(y,x)+R^{\prime(5)}_2(y,x)+\\
&\quad+e^2:\overline{\psi}(y)\beta^\mu\psi(y)\overline{\psi}(x)\beta^\nu\psi(x)::A_\mu(y)A_\nu(x):,\\
\end{aligned}\end{equation}
where
\begin{equation}\label{aprima1r}\begin{aligned}
R^{\prime(1)}_2(x,y)
&=ie^2g_{\mu\nu}:\overline{\psi}(y)\beta^\mu\psi(y)D_0^{(+)}(y-x)\overline{\psi}(x)\beta^\nu\psi(x):,\\
\end{aligned}\end{equation}
\begin{equation}\label{aprima2r}\begin{aligned}
R^{\prime(2)}_2(x,y)
&=ie^2:\overline{\psi}(x)\beta^{\nu}S^{(-)}(x-y)\beta^\mu\psi(y)::A_\mu(y)A_\nu(x):-\\
&\quad-ie^2:\overline{\psi}(y)\beta^{\mu}S^{(+)}(y-x)\beta^\nu\psi(x)::A_\mu(y)A_\nu(x):,\\
\end{aligned}\end{equation}
\begin{equation}\label{aprima3r}\begin{aligned}
R^{\prime(3)}_2(y,x)
&=e^2:A_\mu(y)Tr[\beta^{\mu}S^{(+)}(y-x)\beta^{\nu}S^{(-)}(x-y)]A_\nu(x):,\\
\end{aligned}\end{equation}
\begin{equation}\label{aprima4r}\begin{aligned}
R^{\prime(4)}_2(y,x)
&=-e^2g_{\mu\nu}:\overline{\psi}_c(x)\beta^\nu_{cd}S^{(-)}_{da}(x-y)D_0^{(+)}(y-x)\beta^\mu_{ab}\psi_b(y):+\\
&\quad+e^2g_{\mu\nu}:\overline{\psi}(y)\beta^{\mu}S^{(+)}(y-x)D_0^{(+)}(y-x)\beta^{\nu}\psi(x):,\\
\end{aligned}\end{equation}
\begin{equation}\label{aprima5r}\begin{aligned}
R^{\prime(5)}_2(y,x)
&=ie^2g_{\mu\nu}Tr[\beta^{\mu}S^{(+)}(y-x)\beta^{\nu}S^{(-)}(x-y)]D_0^{(+)}(y-x).\\
\end{aligned}\end{equation}

The causal distribution $D_{2}$ is obtained via the subtraction (\ref{DnSpr}), it will takes the following form
\begin{equation}\label{Dagrupado}
D_2(x,y)=R'_2(x,y)-A'_2(x,y)=D_2^{(1)}+D_2^{(2)}+D_2^{(3)}+D_2^{(4)}+D_2^{(5)},
\end{equation}
where
\begin{equation}\label{D2(1)}
D_2^{(1)}=ie^2g_{\mu\nu}:\overline{\psi}(y)\beta^\mu\psi(y)\left(D_0^{(+)}(y-x)-D_0^{(+)}(x-y)\right)\overline{\psi}(x)\beta^\nu\psi(x):,
\end{equation}
\begin{equation}\label{D2(2)}\begin{aligned}
D_2^{(2)}
&=ie^2:\overline{\psi}(x)\beta^\nu\left(S^{(+)}(x-y)+S^{(-)}(x-y)\right)\beta^\mu\psi(y)::A_\mu(y)A_\nu(x):-\\
&\quad-ie^2:\overline{\psi}(y)\beta^\mu\left(S^{(+)}(y-x)+S^{(-)}(y-x)\right)\beta^\nu\psi(x)::A_\mu(y)A_\nu(x):,\\
\end{aligned}\end{equation}
\begin{equation}\label{D2(3)}\begin{aligned}
D_2^{(3)}
&=e^2Tr[\beta^\mu{}S^{(+)}(y-x)\beta^\nu{}S^{(-)}(x-y)-\beta^\nu{}S^{(+)}(x-y)\beta^\mu{}S^{(-)}(y-x)]\times\\
&\quad\times:A_\mu(y)A_\nu(x):,\\
\end{aligned}\end{equation}
\begin{equation}\label{D2(4)}\begin{aligned}
D_2^{(4)}
&=-e^2g_{\mu\nu}:\overline{\psi}(x)\beta^\nu{}[S^{(-)}(x-y)D_0^{(+)}(y-x)+S^{(+)}(x-y)D_0^{(+)}(x-y)]\beta^\mu\psi(y):+\\
&\quad+e^2g_{\mu\nu}:\overline{\psi}(y)\beta^\mu{}[S^{(+)}(y-x)D_0^{(+)}(y-x)+S^{(-)}(y-x)D_0^{(+)}(x-y)]\beta^\nu\psi(x):,\\
\end{aligned}\end{equation}
\begin{equation}\label{D2(5)}\begin{aligned}
D_2^{(5)}
&=ie^2\beta^\mu{}S^{(+)}(y-x)\beta_\mu{}S^{(-)}(x-y)D_0^{(+)}(y-x)-\\
&\quad-ie^2\beta^\mu{}S^{(+)}(x-y)\beta_\mu{}S^{(-)}(y-x)D_0^{(+)}(x-y).\\
\end{aligned}\end{equation}

We want to emphasize that each term $D_2^{(i)}$ represents different process in the $S$-matrix. 

\section{Moller scattering}\label{S3}

Now we will determine the differential cross section of Moller process that consists in the elastic scattering of two scalar particles $b(p_i)+b(q_i)\rightarrow{b(p_f)+b(q_f)}$, where $p_i$ and $q_i$ are the momentums before the interaction and the other two ones the momentums after. Therefore, the in and out states take the following forms
\begin{equation}\label{estado inicial Moller}
\begin{aligned}
|in_{\text{Moller}}\rangle&=|\Psi_i\rangle\otimes|\Phi_i\rangle=\int{d}^{3}p_{1}{d}^{3}q_{1}\Psi_{i}(\mathbf{p}_1)\Phi_{i}(\mathbf{q}_1)a^{\dag}(\mathbf{p}_1)a^{\dag}(\mathbf{q}_1)|0\rangle,\\
\end{aligned}
\end{equation}
\begin{equation}\label{estado final Moller}
\begin{aligned}
|out_{\text{Moller}}\rangle&=|\Psi_f\rangle\otimes|\Phi_f\rangle=\int{d}^{3}p_{2}{d}^{3}q_{2}\Psi_{f}(\mathbf{p}_2)\Phi_{f}(\mathbf{q}_2)a^{\dag}(\mathbf{p}_2)a^{\dag}(\mathbf{q}_2)|0\rangle,\\
\end{aligned}
\end{equation}
where $\{\Psi_{i}(p_{1}),\Psi_{f}(p_{2}),\Phi_{i}(q_{1}),\Phi_{f}(q_{2})\}$ are wave packets sharply peaked in $\{p_i,p_f,q_i,q_f\}$ respectively.

Taking into account the creation and annihilation operators in the computation of the scattering amplitude $\langle{}out_{\text{Moller}}|S|in_{\text{Moller}}\rangle$, it is not difficult to note that only the term coming from $D_2^{(1)}$ will be not null.

Besides, by using the property $D_{0}^{(+)}(x)=-D_{0}^{(-)}(-x)$ we can rewrite $D_2^{(1)}(x,y)$ in the following form
\begin{equation}\label{D2DKP2}\begin{aligned}
D_2^{(1)}(x,y)
&=-e^2ig_{\mu\nu}:\overline{\psi}(y)\beta^\mu\psi(y)D_0(x-y)\overline{\psi}(x)\beta^\nu\psi(x):,\\
\end{aligned}\end{equation}
where we can see that its \textit{numerical part} is $D_0(x-y)$.

\subsection{Causal splitting of $D_0$}

To continue our study of Moller process, we must perform the \textit{causal splitting procedure} of $D_0(x-y)$. Therefore, as developed in section \ref{CSP}, we will work with the Fourier transform of $D_0(x-y)$ given in (\ref{FD}) and which we rewrite here
\begin{equation}\label{TransformadaD0}
\widehat{D}_0(p)=\frac{i}{2\pi}\delta({p^2})sgn(p^0).
\end{equation}

The power counting function for $\widehat{D}_0(p)$ is determined using (\ref{limite igual a d03}). Thus, we must compute the form of $\widehat{D}_0(\frac{p}{\alpha})$, this is 
\begin{equation}\label{TransformadaD0peta}\begin{aligned}
\widehat{D}_0(\frac{p}{\alpha})
&=\frac{i}{2\pi}\delta({p^2\alpha^{-2}})sgn(p^0\alpha^{-1})\\
&=\frac{i\alpha^2}{2\pi}\delta({p^2})sgn(p^0\alpha^{-1}).\\
\end{aligned}\end{equation}

From (\ref{TransformadaD0peta}), it is not difficult to conclude that with $\rho(\alpha)=\alpha^{-2}$ we obtain a not null quasi-asymptotic
\begin{equation}\label{limite igual a d0 en p}
\lim_{\alpha\rightarrow0}\rho(\alpha)\left\langle\widehat{D}_0(\frac{p}{\alpha}),\widecheck{f}(p)\right\rangle=\left\langle\widehat{D}_0(p),\widecheck{f}(p)\right\rangle\neq0.
\end{equation}

Therefore, regarding (\ref{0rho23}), the \textit{order of singularity} of $\hat{D}_0$ is
\begin{equation}\label{GrauD0final}
\omega[D_0]=-2,
\end{equation}
which means that $\widehat{D}_0(p)$ is a \textit{regular} distribution and its splitting in retarded and advanced part can be done using the Heaviside step function as
\begin{equation}\label{splitD0}\begin{aligned}
D_0(x-y)
&=\theta(x^0-y^0)D_0(x-y)-\theta(y^0-x^0)D_0(x-y)\\
&=D_0^{\text{ret}}(x-y)-D_0^{\text{adv}}(x-y),\\
\end{aligned}\end{equation}
where $D_0^{\text{ret}}(x-y)=\theta(x^0-y^0)D_0(x-y)$ and $D_0^{\text{adv}}(x-y)=\theta(y^0-x^0)D_0(x-y)$.

From (\ref{D2DKP2}) and (\ref{splitD0}), the second order retarded distribution $R_2^{(1)}(x,y)$ is given by
\begin{equation}\label{R2Moller}
R_2^{(1)}(x,y)=-e^2ig_{\mu\nu}:\overline{\psi}(y)\beta^\mu\psi(y)D_0^{\text{ret}}(x-y)\overline{\psi}(x)\beta^\nu\psi(x):.
\end{equation}

As a last step, with the help of the equations (\ref{SOLTnRES}), (\ref{R2Moller}) and (\ref{aprima1r}), we can determine $T_2^{(1)}(x,y)$ in the following form
\begin{equation}\label{T2Moller}\begin{aligned}
T_2^{(1)}(x,y)
&=R_2^{(1)}-R_2^{\prime(1)}\\
&=-e^2ig_{\mu\nu}\beta^\mu_{ab}\beta^\nu_{cd}:\overline{\psi}_a(y)\psi_b(y)\overline{\psi}_c(x)\psi_d(x):D_0^{\text{ret}}\\
&\quad-[e^2\beta^\mu_{ab}\beta^\nu_{cd}:\overline{\psi}_a(y)\psi_b(y)\overline{\psi}_c(x)\psi_d(x):ig_{\mu\nu}D_0^{(+)}(y-x)]\\
&=-e^2ig_{\mu\nu}\beta^\mu_{ab}\beta^\nu_{cd}:\overline{\psi}_a(y)\psi_b(y)\overline{\psi}_c(x)\psi_d(x):D_0^{F}(x-y),\\
\end{aligned}\end{equation}
where $T_2^{(1)}(x,y)$ is the 2-point distribution associated with the Moller process and $D^F_0(x)\equiv{}D_0^{\text{ret}}(x)-D_0^{(-)}(x)$ is the well known Feynman propagator for a massless scalar field.

\subsection{Computation of the differential cross section}

The non-null scattering amplitude $\mathcal{A}^{(Mo)}_{fi}$  will come from the $S$-matrix term
\begin{equation}\label{S(2)}
\begin{aligned}
S^{(1)}(g)&=\frac{1}{2!}\int d^4yd^4xT_2^{(1)}(x,y)g(x)g(y).
\end{aligned}
\end{equation}

Regarding (\ref{estado inicial Moller}) and (\ref{estado final Moller}), and taking the adiabatic limit for the computations, we have $\mathcal{A}^{(Mo)}_{fi}$ as
\begin{equation}\label{SAM}\begin{aligned}
\mathcal{A}^{(Mo)}_{fi}
&=\langle{}out_{Mo}|S^{(1)}|in_{Mo}\rangle\\
&=\int{d}^{3}p_{2}{d}^{3}q_{2}\int{d}^{3}p_{1}{d}^{3}q_{1}\Psi^{*}_{f}(\mathbf{p}_2)\Phi^{*}_{f}(\mathbf{q}_2)\tilde{\mathcal{A}}^{(Mo)}_{if}\Psi_{i}(\mathbf{p}_1)\Phi_{i}(\mathbf{q}_1)\\
\end{aligned}\end{equation}
where
\begin{equation}\label{SMfi}\begin{aligned}
\tilde{\mathcal{A}}^{(Mo)}_{if}
&=\langle0|a(\mathbf{p}_2)a(\mathbf{q}_2)S^{(1)}a^{\dag}(\mathbf{p}_1)a^{\dag}(\mathbf{q}_1)|0\rangle\\
&=-\frac{1}{2!}\int{d}^4yd^4xe^2ig_{\mu\nu}\beta^\mu_{ab}\beta^\nu_{cd}D_0^{F}(x-y)\times\\
&\quad\times\langle0|{a}(\mathbf{p}_2){a}(\mathbf{q}_2)\overline{\psi}^{(+)}_a(y)\overline{\psi}^{(+)}_c(x)
\psi^{(-)}_b(y)\psi^{(-)}_d(x){a}^\dag(\mathbf{p}_1){a}^\dag(\mathbf{q}_1)|0\rangle\\
\end{aligned}\end{equation}

By using Wick theorem, the contractions (\ref{cont1A}) and (\ref{cont2A}), and the formula
\begin{equation}\label{general integral}\begin{aligned}
\int{d^4x}{d^4y}&D^{F}_0(x-y)e^{iAx+iBy}
=(2\pi)^4\delta(A+B)[-\frac{1}{A^2+i0}],\\
\end{aligned}\end{equation}
we can reduce the expression (\ref{SMfi}) into
\begin{equation}\label{SMfi4}\begin{aligned}
\tilde{\mathcal{A}}^{(Mo)}_{fi}=\delta(q_2-q_1+p_2-p_1)\mathcal{M},
\end{aligned}\end{equation}
where
\begin{equation}\label{MMoller}\begin{aligned}
\mathcal{M}
&=\frac{e^2ig_{\mu\nu}}{(2\pi)^2}\Big[
\overline{u^{-}}_a(\mathbf{p}_2)\beta^\mu_{ab}u_b^-(\mathbf{p}_1)\overline{u^{-}}_c(\mathbf{q}_2)\beta^\nu_{cd}u_d^-(\mathbf{q}_1)\frac{1}{(q_2-q_1)^2+i0}\\
&+\overline{u^{-}}_a(\mathbf{p}_2)\beta^\mu_{ab}u_b^-(\mathbf{q}_1)\overline{u^{-}}_c(\mathbf{q}_2)\beta^\nu_{cd}u_d^-(\mathbf{p}_1)\frac{1}{(q_2-p_1)^2+i0}
\Big]\\
\end{aligned}\end{equation}

As is known, in the center-of-mass reference the differential cross section is given by the following formula
\begin{equation}\begin{aligned}\label{sigma centro de masa dif2}
\frac{d\sigma_{c.m}}{d\Omega}&=(2\pi)^2\frac{E^2}{4}|\mathcal{M}|^2.
\end{aligned}\end{equation}

In the particular case of the Moller scattering, we have for the factor $|\mathcal{M}|^{2}$ the following expression
\begin{equation}\label{M2b}\begin{aligned}
&|\mathcal{M}|^2=\\
&=\frac{e^4}{(2\pi)^4}\Bigg[
\frac{1}{16m^4{p}^0_f{p}^0_i{q}^0_f{q}^0_i}\frac{g_{\mu\alpha}g_{\nu\omega}}{(q_f-q_i)^4}
Tr\bigg\{\slashed{p}_f(\slashed{p}_f+m)\beta^{\alpha}\slashed{p}_i(\slashed{p}_i+m)\beta^{\omega}\bigg\}
Tr\bigg\{\slashed{q}_f(\slashed{q}_f+m)\beta^{\mu}\slashed{q}_i(\slashed{q}_i+m)\beta^\nu\bigg\}
\\
&\quad+
\frac{1}{16m^4{p}^0_f{p}^0_i{q}^0_f{q}^0_i}\frac{g_{\mu\alpha}g_{\nu\omega}}{(q_f-q_i)^2(q_f-p_i)^2}
Tr\bigg\{\slashed{p}_f(\slashed{p}_f+m)\beta^{\alpha}\slashed{p}_i(\slashed{p}_i+m)\beta^\nu
\slashed{q}_f(\slashed{q}_f+m)\beta^\mu\slashed{q}_i(\slashed{q}_i+m)\beta^\omega\bigg\}
\\
&\quad+
\frac{1}{16m^4{p}^0_f{q}^0_i{q}^0_f{p}^0_i}\frac{g_{\mu\alpha}g_{\nu\omega}}{(q_f-p_i)^2(q_f-q_i)^2}
Tr\bigg\{\slashed{p}_f(\slashed{p}_f+m)]\beta^\alpha\slashed{q}_i(\slashed{q}_i+m)\beta^\nu
\slashed{q}_f(\slashed{q}_f+m)\beta^\mu\slashed{p}_i(\slashed{p}_i+m)\beta^{\omega}\bigg\}
\\
&\quad+
\frac{1}{16m^4{p}^0_f{q}^0_i{q}^0_f{p}^0_i}\frac{g_{\mu\alpha}g_{\nu\omega}}{(q_f-p_i)^4}
Tr\bigg\{\slashed{p}_f(\slashed{p}_f+m)\beta^\alpha\slashed{q}_i(\slashed{q}_i+m)\beta^\omega\bigg\}
Tr\bigg\{\slashed{q}_f(\slashed{q}_f+m)\beta^\mu\slashed{p}_i(\slashed{p}_i+m)\beta^\nu\bigg\}
\Bigg].\\
\end{aligned}\end{equation}

The traces can be performed with the help of properties (\ref{TRAZ1}) and (\ref{TRAZ2}). In the bellow expression we present its final result
\begin{equation}\label{m222}\begin{aligned}
|\mathcal{M}|^2&=\frac{e^4}{(2\pi)^4}\frac{1}{4E^4}\left|\frac{(p_iq_i)+(q_fq_i)}{(q_f-p_i)^2}+\frac{(q_ip_i)+(p_fq_i)}{(p_f-p_i)^2}\right|^2.\\
\end{aligned}\end{equation}

In order to compare with other results, we will use the Mandelstam variables $s=(p_i+q_i)^2=(p_f+q_f)^2$, $t=(p_i-p_f)^2$, $u=(p_i-q_f)^2$, to rewrite the expression (\ref{m222}) in the following form
\begin{equation}\label{m prima quadrado}
|\mathcal{M}|^{2}=\frac{e^4}{(2\pi)^4}\frac{1}{16E^4}\left|\frac{s-t}{u}+\frac{s-u}{t}\right|^2.
\end{equation}

Replacing (\ref{m prima quadrado}) into (\ref{sigma centro de masa dif2}), we finally obtain
\begin{equation}\begin{aligned}\label{sigma centro de masa dif B}
\frac{d\sigma_{c.m}}{d\Omega}&=\frac{\alpha^2}{4s}\left|\frac{s-t}{u}+\frac{s-u}{t}\right|^2.
\end{aligned}\end{equation}

The final result (\ref{sigma centro de masa dif2}), is identical to that obtained by C. Itzykson and J. B. Zuber in \cite{Itz} and by J. Beltran in \cite{Jhosep}.

\section{Compton scattering}\label{S4}

The Compton scattering process is represented as
\begin{equation}\label{procesos compton como reaccion}
b(p_i)+\gamma(k_i)\rightarrow{b}(p_f)+\gamma(k_f),
\end{equation}
where $b(p_{i,f})$ represents a scalar with momentum $p_{i,f}$ and $\gamma(k_{i,f})$ a photon with momentum $k_{i,f}$.

Therefore, the in and out states can be written as follows
\begin{equation}\label{estado inicial compton}
\begin{aligned}
|in_{\text{Comp}}\rangle&=|\Psi_i\rangle\otimes|\Phi_i\rangle\\
&=\int{d}^{3}p_{1}{d}^{3}k_{1}\Psi_{i}(\mathbf{p}_1)\Phi_{i}(\mathbf{k}_1){a}^\dag(\mathbf{p}_1)\varepsilon_{i\nu}(\mathbf{k}_1){c}_{\nu}^\dag(\mathbf{k}_1)|0\rangle,
\end{aligned}
\end{equation}
\begin{equation}\label{estado final compton}
\begin{aligned}
|out_{\text{Comp}}\rangle&=|\Psi_f\rangle\otimes|\Phi_f\rangle\\
&=\int{d}^{3}p_{2}{d}^{3}k_{2}\Psi_{f}(\mathbf{p}_2)\Phi_{f}(\mathbf{k}_2){a}^\dag(\mathbf{p}_2)\varepsilon_{f\mu}(\mathbf{k}_2){c}_{\mu}^\dag(\mathbf{k}_2)|0\rangle,
\end{aligned}
\end{equation}
where  $\Psi_{i,f}(\mathbf{p}_1)$ and $\Phi_{i,f}(\mathbf{k}_2)$ are wave packets sharply peaked in $\mathbf{p}_{i,f}$ and $\mathbf{k}_{i,f}$, ${a}^\dag$ and ${c}_{\nu}^\dag$ are the creation operators of a scalar particle and a photon, and  $\varepsilon_{i\nu}$ and $\varepsilon_{f\mu}$ are the initial and final vector polarization for photons, respectively.

Therefore, the transition scattering amplitude $\mathcal{A}_{fi}=\langle{out}_{\text{Comp}}|S|in_{\text{Comp}}\rangle$ is expressed as follows
\begin{equation}\label{TMC}
\mathcal{A}_{fi}^{Comp}=\int{d}^{3}p_{2}{d}^{3}k_{2}\int{d}^{3}p_{1}{d}^{3}k_{1}\Psi^{*}_{f}(\mathbf{p}_2)\Phi^{*}_{f}(\mathbf{k}_2)\tilde{\mathcal{A}}_{fi}^{Comp}\Psi_{i}(\mathbf{p}_1)\Phi_{i}(\mathbf{k}_1),
\end{equation}
where
\begin{equation}\label{TMCq}
\tilde{\mathcal{A}}_{fi}^{Comp}=\langle0|{a}(\mathbf{p}_2)\varepsilon_{f\mu}(\mathbf{k}_2){c}_{\mu}(\mathbf{k}_2){S}{a}^\dag(\mathbf{p}_1)\varepsilon_{i\nu}(\mathbf{k}_1){c}_{\nu}^\dag(\mathbf{k}_1)|0\rangle.
\end{equation}

From (\ref{TMCq}), it is not difficult to note that the term of $T_{2}$, which give us a non-null value, will come from $D_2^{(2)}$. In consequence, we must causal split these term of $D_2$. we can rewrite $D_2^{(2)}$ here in the following form
\begin{equation}\label{D2(2)3}\begin{aligned}
D_2^{(2)}=
&e^2i:\overline{\psi}(x)\beta^\nu{}S(x-y)\beta^\mu\psi(y)::A_\mu(y)A_\nu(x):-\\
&-e^2i:\overline{\psi}(y)\beta^\mu{}S(y-x)\beta^\nu\psi(x)::A_\mu(y)A_\nu(x):,\\
\end{aligned}\end{equation}
where we can see that the numerical parts to causal split are $S(x-y)$ and $S(y-x)$.

\subsection{The causal splitting of $S(x-y)$}

In momentum space the function $\hat{S}(p)$ had given in (\ref{SFT}). Thus, in order to determine the order of singularity $\omega$ of $\hat{S}(p)$, we will compute the form of $\hat{S}(p/\alpha)$
\begin{equation}\label{S1TFeta}\begin{aligned}
\hat{S}(\frac{p}{\alpha})
&=\frac{i}{2\pi{}m}[\slashed{p}\alpha^{-1}(\slashed{p}\alpha^{-1}+m)]\delta(p^2\alpha^{-2}-m^2)sgn(p^0\alpha^{-1})\\
&=\frac{i}{2\pi{}m}[\slashed{p}(\slashed{p}+m\alpha)]\delta(p^2-\alpha^2m^2)sgn(p^0\alpha^{-1}).\\
\end{aligned}\end{equation}

Now, by using (\ref{S1TFeta}) and (\ref{limite igual a d03}), we can see that for a power counting function $\rho(\eta)=1$, we obtain the following non-null quasi-asymptotic distribution
\begin{equation}\label{LimiteS}
\lim_{\alpha\rightarrow0}\rho(\alpha)\langle{}\hat{S}(\frac{p}{\alpha}),\check{f}(p)\rangle=
\langle{}\frac{i}{2\pi{}m}[\slashed{p}\slashed{p}]\delta(p^2)sgn(p^0),\check{f}(p)\rangle\neq0.
\end{equation}

Therefore, from (\ref{0rho23}), we can obtain the following order of singularity
\begin{equation}\label{omega compton}
\omega[\hat{S}(p)]=0.
\end{equation}
Differently to the Moller process, here we obtain that $\hat{S}(p)$ is a singular distribution. This result agree with the order of singularity founded by G. Scharf et al. \cite{Scharf28} for the same process but using Klein-Gordon-Fock formalism. Furtheremore, is in contradiction with the standard formalism of QFT where all tree level processes are considered regular ones.  

In consequence of its singular nature, the retarded part of $\hat{S}(p)$ is given by (\ref{retfmen12No11}). By Replacing (\ref{SFT}) into (\ref{retfmen12No11}), we have
\begin{equation}\label{rsingforS2}\begin{aligned}
\hat{r}_{0}(p)
&=\frac{1}{m}[\slashed{p}(\slashed{p}+m)]\{\frac{i}{2\pi}sgn(p^0)\int_{0}^{\infty}dt\frac{2t\widehat{D}_m(pt)}{(1-t+sgn(p^0)i0^+)}\}+C,\\
\end{aligned}\end{equation}
where the constant $C$ is a $5\times5$ matrix which is not fixed by the causal splitting procedure. 

For the sake of evaluating the integral in (\ref{rsingforS2}), we can use the fact that the order of singularity of $\hat{D}_{m}(p)$ is $\omega[\hat{D}_{m}(p)]=-2$. Thus, using (\ref{rn23}), it is not difficult to note that the factor between braces in (\ref{rsingforS2}) is $D_m^{\text{ret}}(p)$. Therefore, the most general solution for the retarded part $\tilde{S}^{\text{ret}}(p)$ is equal to
\begin{equation}\label{FD4}
\tilde{S}^{\text{ret}}(p)=\frac{1}{m}[\slashed{p}(\slashed{p}+m)]D_m^{\text{ret}}(p)+C.
\end{equation}

Furthermore, in the same style of equations (\ref{S+-FT}) or (\ref{SFT}), we can write the first term of the right hand side of equation (\ref{FD4}) as follows
\begin{equation}\label{FD4I}
{S}^{\text{ret}}(p)=\frac{1}{m}[\slashed{p}(\slashed{p}+m)]D_m^{\text{ret}}(p),
\end{equation}
which is the retarded part obtained via the product of ${S}(x-y)$ by the Heaviside step function. The latter means that we could split ${S}(x-y)$ by the standard procedure as
\begin{equation}\label{SS}
S(x-y)=S^{ret}(x-y)-S^{adv}(x-y),
\end{equation}
but, \textbf{\textit{CPT tell us that this splitting is not unique}} because the singular order of ${S}(x-y)$. In configuration space, the most general solution (\ref{FD4}) will be rewritten as
\begin{equation}\label{GRS}
\tilde{S}^{ret}(x-y)=S^{ret}(x-y)+C\delta(x-y).
\end{equation}

We will use the configuration space solution (\ref{GRS}) to fix $C$ later.

On the other hand, by noting that the numerical part $S(y-x)$ has opposite sign in its input part, we conclude that its retarded par is given by
\begin{equation}\label{RETS}
\tilde{S}^{ret}(y-x)=-S^{adv}(y-x)+C'\delta(x-y)
\end{equation}

With the help of (\ref{GRS}) and (\ref{RETS}), the retarded distribution $R_2^{(2)}$ is
\begin{equation}\label{D2(2)$}\begin{aligned}
R_2^{(2)}=
&e^2i:\overline{\psi}(x)\beta^\nu\Big(S^{ret}(x-y)+C\delta(x-y)\Big)\beta^\mu\psi(y)::A_\mu(y)A_\nu(x):-\\
&-e^2i:\overline{\psi}(y)\beta^\mu(-S^{adv}(y-x)+C\delta(x-y))\beta^\nu\psi(x)::A_\mu(y)A_\nu(x):.\\
\end{aligned}\end{equation}

Now, by performing the difference $T_2^{(2)}=R_2^{(2)}-R_2^{\prime(2)}$, we obtain for the two point distribution, associated with the Compton process, the following form
\begin{equation}\label{T2(2)63}\begin{aligned}
T_2^{(2)}(x,y)
&=e^2i:\overline{\psi}(x)\beta^\nu\Big(-S^{F}(x-y)+C\delta(x-y)\Big)\beta^\mu\psi(y)::A_\mu(y)A_\nu(x):+\\
&\quad+e^2i:\overline{\psi}(y)\beta^\mu\Big(-S^{F}(y-x)-C^{\prime}\delta(x-y)\Big)\beta^\nu\psi(x)::A_\mu(y)A_\nu(x):,\\
\end{aligned}\end{equation}
where $S^{F}(x)=S^{(-)}(x)-S^{ret}(x)=-S^{(+)}(x)-S^{adv}(x)$ is the feynman propagator as usual.

Because of the symmetry  property of $T_2^{(2)}(x,y)$ under the interchange of variable $x\rightleftarrows{}y$, we can see that $C'=-C$, and obtain
\begin{equation}\label{T2(2)64}\begin{aligned}
T_2^{(2)}(x,y)
&=e^2i:\overline{\psi}(x)\beta^\nu\Big(-S^{F}(x-y)+C\delta(x-y)\Big)\beta^\mu\psi(y)::A_\mu(y)A_\nu(x):+\\
&\quad+e^2i:\overline{\psi}(y)\beta^\mu\Big(-S^{F}(y-x)+C\delta(x-y)\Big)\beta^\nu\psi(x)::A_\mu(y)A_\nu(x):.\\
\end{aligned}\end{equation}

\subsection{Fixation of constants $C$}

Because of the singular nature of $D^{(2)}_{2}(x,y)$, the \textit{causal splitting procedure} (based on \textit{causality} and \textit{gauge invariance} at first order) give us a family of 2-point causal distributions $T_2^{(2)}(x,y)$ represented in the freedom of constant $C$. To fix this constant, we must use other physical properties of the theory.

Graphically, the Compton scattering has two external photon legs, this allows us to use \textbf{\textit{perturbative gauge invariance}} at second order to determine $C$. Therefore, we need to compute the gauge derivative $d_{Q}T_{2}(x,y)$, using (\ref{GT1}) this result is
\begin{equation}\label{TnP2}\begin{aligned}
d_{Q}T_{2}(x,y)
&=e^2i:\overline{\psi}(x)\beta^\nu\Big(-S^{F}(x-y)+C\delta(x-y)\Big)\beta^\mu\psi(y)::i\partial_{\mu}u(y)A_\nu(x):+\\
&\quad+e^2i:\overline{\psi}(x)\beta^\nu\Big(-S^{F}(x-y)+C\delta(x-y)\Big)\beta^\mu\psi(y)::A_\mu(y)i\partial_{\nu}u(x):+\\
&\quad+e^2i:\overline{\psi}(y)\beta^\mu\Big(-S^{F}(y-x)+C\delta(x-y)\Big)\beta^\nu\psi(x)::A_\mu(y)i\partial_{\nu}u(x):+\\
&\quad+e^2i:\overline{\psi}(y)\beta^\mu\Big(-S^{F}(y-x)+C\delta(x-y)\Big)\beta^\nu\psi(x)::i\partial_{\mu}u(y)A_\nu(x):\\
\end{aligned}\end{equation}

Defining $Q^{\nu\mu}_{xy}$ as
\begin{equation}\label{T2}
\begin{aligned}
Q^{\nu\mu}_{xy}
&=:\overline{\psi}(x)\beta^\nu\Big[-S^{F}(x-y)+C\delta(x-y)\Big]\beta^\mu\psi(y):+\\
&\quad+:\overline{\psi}(y)\beta^\mu\Big[-S^{F}(y-x)+C\delta(x-y)\Big]\beta^\nu\psi(x):,\\
\end{aligned}
\end{equation}
we can rewrite (\ref{TnP2}), as
\begin{equation}\label{TnP23}\begin{aligned}
d_{Q}T_{2}(x,y)
&=e^2i^{2}\partial_{\mu}^{y}\left(Q^{\nu\mu}_{xy}:u(y)A_\nu(x):\right)-e^2i^{2}\partial_{\mu}^{y}(Q^{\nu\mu}_{xy}):u(y)A_\nu(x):+\\
&\quad+e^2i^{2}\partial_{\nu}^{x}\left(Q^{\nu\mu}_{xy}:A_\mu(y)u(x):\right)-e^2i^{2}\partial_{\nu}^{x}(Q^{\nu\mu}_{xy}):A_\mu(y)u(x):.\\
\end{aligned}\end{equation}

Examining (\ref{TnP}) and (\ref{TnP23}), it is clear that to get a gauge invariance $S$-matrix at second order, it is necessary to fulfill the following conditions
\begin{equation}\label{2GIC}
\partial_{\nu}^{x}(Q^{\nu\mu}_{xy})=0=\partial_{\mu}^{y}(Q^{\nu\mu}_{xy}).
\end{equation}

By evaluating the first derivative of (\ref{2GIC}), we obtain
\begin{equation}\label{derivando respecto de x llaves}
\begin{aligned}
\partial_{\nu,x}{}Q^{\mu\nu}
&=:\partial_{\nu,x}\overline{\psi}(x)\beta^\nu[-S^{F}(x-y)+C\delta(x-y)]\beta^\mu\psi(y):+\\
&\quad+:\overline{\psi}(x)\beta^\nu[-\partial_{\nu,x}S^{F}(x-y)+C\partial_{\nu,x}\delta(x-y)]\beta^\mu\psi(y):+\\
&\quad+:\overline{\psi}(y)\beta^\mu[-\partial_{\nu,x}S^{F}(y-x)+C\partial_{\nu,x}\delta(x-y)]\beta^\nu\psi(x):+\\
&\quad+:\overline{\psi}(y)\beta^\mu[-S^{F}(y-x)+C\delta(x-y)]\beta^\nu\partial_{\nu,x}\psi(x):.\\
\end{aligned}
\end{equation}

Besides, we can write the DKP Feynman propagator $S^{F}(x)$ regarding (\ref{JPS}) and (\ref{FD4}), thus we obtain
\begin{equation}\label{SFeynman}
S^{F}(x)=S^{(-)}(x)-S^{ret}(x)=-S^{(+)}(x)-S^{adv}(x)=-\frac{1}{m}[i\slashed{\partial}(i\slashed{\partial}+m)]D^{F}(x).
\end{equation}

In order to obtain the derivative of $S^{F}(x)$, that we need to replace in (\ref{derivando respecto de x llaves}), we can multiplied (\ref{SFeynman}) from the left by $(i\slashed{\partial}-m)$ to obtain the following expression
\begin{equation}\label{SFeynmanprop1}\begin{aligned}
i\slashed{\partial}S^{F}(x)&=mS^{F}(x)+\frac{i}{m}\slashed{\partial}\delta(x).\\
\end{aligned}\end{equation}

On the other hand, regarding the property $\beta^{\mu\dag}=\eta^{0}\beta^{\mu}\eta^{0}$, we can demonstrate that the transpose conjugate of $S^{F}(x)$ is given by $(S^{F}(x))^\dag=\eta^0S^{F}(-x)\eta^0$. After this, we can conjugate (\ref{SFeynmanprop1}) to get the following result
\begin{equation}\label{SFeynmanprop2}\begin{aligned}
-i{\partial}_{\nu}S^{F}(-x)\beta^{\nu}
&=mS^{F}(-x)-\frac{i}{m}\slashed{\partial}\delta(x).\\
\end{aligned}\end{equation}

Replacing (\ref{SFeynmanprop1}) and (\ref{SFeynmanprop2}) into (\ref{derivando respecto de x llaves}), we have
\begin{equation}\label{derivando respecto de x llaves2}
\begin{aligned}
\partial_{\nu,x}{}Q^{\mu\nu}
&=+:\overline{\psi}(x)[-\frac{1}{m}\slashed{\partial}\delta(x)+C\slashed{\partial}\delta(x-y)]\beta^\mu\psi(y):+\\
&\quad+:\overline{\psi}(y)\beta^\mu[-\frac{1}{m}\slashed{\partial}\delta(x)+C\slashed{\partial}\delta(x-y)]\psi(x):,\\
\end{aligned}
\end{equation}
where we can see clear that to satisfy the condition (\ref{2GIC}), $C$ must be
\begin{equation}\label{CCCCCCC}
C=\frac{I}{m},
\end{equation}
where $I$ is the $5\times5$ identity matrix.

\subsection{Computation of the differential cross section}

By replacing ($\ref{CCCCCCC}$) into (\ref{T2(2)64}), we obtain the $2$-point distribution associated with the Compton process $T_2^{(2)}(x,y)$ given by the following expression
\begin{equation}\label{T2(2)65}\begin{aligned}
T_2^{(2)}(x,y)
&=e^2i:\overline{\psi}(x)\beta^\nu\left(-S^{F}(x-y)+\frac{I}{m}\delta(x-y)\right)\beta^\mu\psi(y)::A_\mu(y)A_\nu(x):\\
&\quad+e^2i:\overline{\psi}(y)\beta^\mu\left(-S^{F}(y-x)+\frac{I}{m}\delta(x-y)\right)\beta^\nu\psi(x)::A_\mu(y)A_\nu(x):.\\
\end{aligned}\end{equation}

In order to compute the differential cross section, we will denote as $S^{(2)}_2$ the term of $S$-matrix associated with the Compton scattering. Regarding (\ref{T2(2)65}), $S^{(2)}_2$ can be written in the following form
\begin{equation}\label{S(2)x}\begin{aligned}
S^{(2)}_2
&=\frac{1}{2}\int{d^4x}{d^4y}T_2^{(2)}(x,y)g(x)g(y)\\
&=-\frac{1}{2}\int{d^4x}{d^4y}e^2i:\overline{\psi}(x)\beta^\nu_{cd}S^{F}(x-y)\beta^\mu\psi(y)::A_\mu(y)A_\nu(x):g(x)g(y)-\\
&\quad-\frac{1}{2}\int{d^4x}{d^4y}e^2i:\overline{\psi}(y)\beta^\mu_{ab}S^{F}(y-x)\beta^\nu\psi(x)::A_\mu(y)A_\nu(x):g(x)g(y)+\\
&\quad+\frac{e^2i}{m}\int{d^4x}:\overline{\psi}(x)\beta^\mu\beta^\nu\psi(x)::A_\mu(x)A_\nu(x):g^2(x),\\
\end{aligned}\end{equation}
where in the last integral we had joined the two terms coming from the Dirac delta functions. \textbf{\textit{This term could be seen as a graph where we have two photons and two scalars legs interacting in the same point}}. As pointed out by Akhiezer and Berestetskii in \cite{Akh}, an advantage of DKP theory is that this term do not appear. But, from (\ref{S(2)x}), it is indisputable that such term appears because of the singular nature of the causal propagator associated. The same happens in SQED$_4$ when it is studied by CPT too \cite{Scharf28}.

To continue, we will split $S^{(2)}_2$ in the following form
\begin{equation}\label{Scomptres}
S^{(2)}_2=_aS^{(2)}_2+_bS^{(2)}_2,
\end{equation}
where
\begin{equation}\label{aS22comp}
_aS^{(2)}_2
=\frac{e^2i}{m}\int{d^4x}:\overline{\psi}(x)\beta^\mu\beta^\nu\psi_d(x)::A_\mu(x)A_\nu(x):g^2(x),
\end{equation}
\begin{equation}\label{bS22comp}\begin{aligned}
_bS^{(2)}_2
&=-\frac{e^2i}{2}\int{d^4x}{d^4y}:\overline{\psi}(x)\beta^\nu{}S^{F}(x-y)\beta^\mu\psi(y)::A_\mu(y)A_\nu(x):g(x)g(x)\\
&\quad-\frac{e^2i}{2}\int{d^4x}{d^4y}:\overline{\psi}(y)\beta^\mu{}S^{F}(y-x)\beta^\nu\psi_d(x)::A_\mu(y)A_\nu(x):g(x)g(x).\\
\end{aligned}\end{equation}

Therefore, the scattering amplitude distribution $\tilde{S}_{fi}^{Comp}$ can be written as
\begin{equation}\label{SifCI}
\tilde{\mathcal{A}}_{fi}^{Comp}=_a\mathcal{A}^{(2)}_{if}+_b\mathcal{A}^{(2)}_{if}
\end{equation}
where, in the adiabatic limit $g(x)\rightarrow1$, we have
\begin{equation}\label{S(2)1}\begin{aligned}
_a\mathcal{A}^{(2)}_{if}
&=\frac{e^2i}{m}\int{d^4x}\langle0|a(\mathbf{p}_2):\overline{\psi}(x)\beta^\mu\beta^\nu\psi(x):a^\dag(\mathbf{p}_1)|0\rangle\times\\
&\quad\times\langle0|\varepsilon_{f\beta}(\mathbf{k}_2){c}_{\beta}(\mathbf{k}_2):A_\mu(x)A_\nu(x):\varepsilon_{i\alpha}(\mathbf{k}_1){c}_{\alpha}^\dag(\mathbf{k}_1)|0\rangle,\\
\end{aligned}\end{equation}
\begin{equation}\label{S(2)1B}\begin{aligned}
_b\mathcal{A}^{(2)}_{if}
&=-e^2i\int{d^4x}{d^4y}\langle0|a(\mathbf{p}_2):\overline{\psi}(x)\beta^\nu{}S^{F}(x-y)\beta^\mu\psi(y):a^\dag(\mathbf{p}_1)|0\rangle\times\\
&\quad\times\langle0|\varepsilon_{f\beta}(\mathbf{k}_2){c}_{\beta}(\mathbf{k}_2):A_\mu(y)A_\nu(x):\varepsilon_{i\alpha}(\mathbf{k}_1){c}_{\alpha}^\dag(\mathbf{k}_1)|0\rangle.\\
\end{aligned}\end{equation}

Before to reduce the expressions (\ref{S(2)1}) and (\ref{S(2)1B}), firts of all we will consider a real polarization vector $\varepsilon_{\nu}$ with the following properties
\begin{equation}\label{E}
\varepsilon_{\nu}=(0,\bm{\varepsilon}),\quad \bm{\varepsilon}.\bm{k}=0,\quad \bm{\varepsilon}^{2}=1.
\end{equation}

Secondly, we will use the following contractions between the electromagnetic field $A^{\mu}(x)$ and the creation or annihilation operator for photons
\begin{equation}\label{PC1}\begin{aligned}
\contraction{\varepsilon_{f\beta}(\mathbf{k}_j)}{{c}}{_{\beta}(\mathbf{k}_j)}{A}\varepsilon_{f\beta}(\mathbf{k}_j){c}_{\beta}(\mathbf{k}_j)A_\mu(x)
&=(2\pi)^{-3/2}\dfrac{\varepsilon_{f\mu}(\mathbf{k}_j)}{\sqrt{2\omega_j}}e^{ik_jx},
\end{aligned}\end{equation}
\begin{equation}\label{PC2}\begin{aligned}
\contraction{}{A}{_{\mu}(x)\varepsilon_{i\beta}(\mathbf{k}_{j})}{c}
A_{\mu}(x)\varepsilon_{i\beta}(\mathbf{k}_{j})c_{\beta}^{\dag}(\mathbf{k}_{j})
&=(2\pi)^{-3/2}\dfrac{\varepsilon_{i\mu}(\mathbf{k}_{j})}{\sqrt{2\omega}}e^{-ik_jx}.\\
\end{aligned}\end{equation}

Returning to the computation of $_a\mathcal{A}^{(2)}_{if}$, we can use Wick theorem and contractions (\ref{cont1A}), (\ref{cont2A}), (\ref{PC1}) and (\ref{PC2}) to reduce the vacuum expectation values in (\ref{S(2)1}), this is
\begin{equation}\label{comp1s}\begin{aligned}
\langle0|a(\mathbf{p}_2)&:\overline{\psi}(x)\beta^\mu\beta^\nu\psi(x):a^\dag(\mathbf{p}_1)|0\rangle=\\
&=
\contraction{\langle0|}{a}{(\mathbf{p}_f):}{\bar{\psi}}
\contraction{\langle0|a(\mathbf{p}_f):\bar{\psi}_a(x)\beta^\mu_{ac}\beta^\nu_{cd}}{\psi}{_d(x):}{a}
\langle0|a(\mathbf{p}_2):\bar{\psi}_a(x)\beta^\mu_{ac}\beta^\nu_{cd}\psi_d(x):a^\dag(\mathbf{p}_1)|0\rangle\\
&=\frac{1}{(2\pi)^{3}}\overline{u^{-}}(\mathbf{p}_2)\beta^\mu\beta^\nu{}u^-(\mathbf{p}_1)e^{-i(p_1-p_2)x},\\
\end{aligned}\end{equation}
\begin{equation}\label{acontcomp}\begin{aligned}
\langle0|&\varepsilon_{f\beta}(\mathbf{k}_2){a}_{\beta}(\mathbf{k}_2):A_\mu(x)A_\nu(x):\varepsilon_{i\alpha}(\mathbf{k}_1){a}_{\alpha}^\dag(\mathbf{k}_1)|0\rangle=\\
&=
\contraction{\langle0|\varepsilon_{f\beta}(\mathbf{k}_f)}{{a}}{_{\mu}(\mathbf{k}_f):}{A}
\contraction{\langle0|\varepsilon_{f\beta}(\mathbf{k}_f){a}_{\beta}(\mathbf{k}_f):A_\mu(x)}{A}{_\nu(x):\varepsilon_{i\alpha}(\mathbf{k}_i)}{{a}}
\langle0|\varepsilon_{f\beta}(\mathbf{k}_f){a}_{\beta}(\mathbf{k}_f):A_\mu(x)A_\nu(x):\varepsilon_{i\alpha}(\mathbf{k}_i){a}_{\alpha}^\dag(\mathbf{k}_i)|0\rangle+\\
&
\contraction{\quad+\langle0|\varepsilon_{f\beta}(\mathbf{k}_2)}{{a}}{_{\beta}(\mathbf{k}_2):A_\mu(x)}{A}
\bcontraction{\quad+\langle0|\varepsilon_{f\beta}(\mathbf{k}_2){a}_{\beta}(\mathbf{k}_2):}{A}{_\mu(x)A_\nu(x):\varepsilon_{i\alpha}(\mathbf{k}_1)}{{a}}
\quad+\langle0|\varepsilon_{f\beta}(\mathbf{k}_2){a}_{\beta}(\mathbf{k}_2):A_\mu(x)A_\nu(x):\varepsilon_{i\alpha}(\mathbf{k}_1){a}_{\alpha}^\dag(\mathbf{k}_1)|0\rangle\\
&=(2\pi)^{-3}\frac{\varepsilon_{f\mu}(\mathbf{k}_2)}{\sqrt{2\omega_{f}}}\frac{\varepsilon_{i\nu}(\mathbf{k}_1)}{\sqrt{2\omega_{i}}}e^{-i(k_1-k_2)x}
+(2\pi)^{-3}\frac{\varepsilon_{f\nu}(\mathbf{k}_2)}{\sqrt{2\omega_{f}}}\frac{\varepsilon_{i\mu}(\mathbf{k}_1)}{\sqrt{2\omega_{i}}}e^{-i(k_1-k_2)x}.\\
\end{aligned}\end{equation}

Replacing (\ref{comp1s}) and (\ref{acontcomp}) into (\ref{S(2)1}), we obtain
\begin{equation}\label{S(2)2}\begin{aligned}
_a\mathcal{A}^{(2)}_{if}
&=\delta(p_1-p_2+k_1-k_2)\mathcal{M}_{a},
\end{aligned}\end{equation}
where
\begin{equation}\label{Mcomp1}\begin{aligned}
\mathcal{M}_a
&=\frac{ie^2}{m(2\pi)^{2}\sqrt{2\omega_{1}}\sqrt{2\omega_{2}}}
[\overline{u^{-}}(\mathbf{p}_2)\slashed{\varepsilon}_{f}\slashed{\varepsilon}_{i}{}u^-(\mathbf{p}_1)+\overline{u^{-}}(\mathbf{p}_2)\slashed{\varepsilon}_{i}\slashed{\varepsilon}_{f}{}u^-(\mathbf{p}_1)].\\
\end{aligned}\end{equation}

For the reduction of $_b\mathcal{A}^{(2)}_{if}$, the computation is similar but not equal because of $S(x-y)$ function between the $\beta$-matrices. The final result for $_b\mathcal{A}^{(2)}_{if}$ is
\begin{equation}\label{aS(2)12}\begin{aligned}
_{b}\mathcal{A}^{(2)}_{if}
&=\delta(p_2+k_2-p_1-k_1)\mathcal{M}_{b},
\end{aligned}\end{equation}
where
\begin{equation}\label{McompFy}\begin{aligned}
\mathcal{M}_{b}
&=-\frac{e^2i}{m(2\pi)^{2}\sqrt{2\omega_{f}}\sqrt{2\omega_{1}}}\Bigg[\frac{\overline{u^{-}}(\mathbf{p}_2)\slashed{\varepsilon}_{i}(\slashed{p}_1-\slashed{k}_2)(\slashed{p}_1-\slashed{k}_2+m)\slashed{\varepsilon}_{f}{}u^-(\mathbf{p}_1)}{({p}_1-{k}_2)^2-m^2}+\\
&\quad+\frac{\overline{u^{-}}(\mathbf{p}_2)\slashed{\varepsilon}_{f}(\slashed{p}_1+\slashed{k}_1)(\slashed{p}_1+\slashed{k}_1+m)\slashed{\varepsilon}_{i}{}u^-(\mathbf{p}_1)}{({p}_1+{k}_1)^2-m^2}\Bigg].\\
\end{aligned}\end{equation}

With (\ref{aS(2)12}) and (\ref{S(2)2}), the transition amplitude $\tilde{\mathcal{A}}_{fi}^{Comp}$ could be written as
\begin{equation}\label{SifC}
\tilde{\mathcal{A}}_{fi}^{Comp}=\delta(p_2+k_2-p_1-k_1)\mathcal{M},\quad \mathcal{M}=\mathcal{M}_{a}+\mathcal{M}_{b}
\end{equation}

Using the laboratory reference frame, the differential cross section is given by
\begin{equation}\label{seccion de choque diferencial comp}\begin{aligned}
\frac{d\sigma}{d\Omega}\Big|_{\text{lab}}&=(2\pi)^{2}\frac{\omega_f^3E_f}{m\omega_i}|\mathcal{M}(p_i,k_i,p_f,k_f)|^2.\\
\end{aligned}\end{equation}
where 
\begin{equation}\label{M}
|\mathcal{M}|^{2}=|\mathcal{M}_{a}|^{2}+\mathcal{M}^{*}_{a}\mathcal{M}_{b}+\mathcal{M}^{*}_{b}\mathcal{M}_{a}+|\mathcal{M}_{b}|^{2}
\end{equation}
and, because the sharply peaked form of the wave packets, we have the following change of variables
\begin{equation}\label{CB}
p_1\rightarrow{}p_i,\quad p_2\rightarrow{}p_f,\quad k_1\rightarrow{}k_i,\quad k_2\rightarrow{}k_f
\end{equation}

Before of the computation of the terms in (\ref{M}), we must take into account two facts. Firstly, because we are working in the laboratory system, then $p_i=(m,\mathbf{0})$. The latter implies that
\begin{equation}\label{pi epsilon}
p_i\varepsilon_i=0, \quad p_i\varepsilon_f=0,
\end{equation}
which complement the polarization conditions (\ref{E}) which has the following covariant form
\begin{equation}\label{k epsilon}
\varepsilon_ik_i=0,\quad \varepsilon_fk_f=0.
\end{equation}

Furthermore, the denominators of fractions in the brackets of (\ref{McompFy}), will be reduced to 
\begin{equation}\label{DR1}
(p_i-k_f)^{2}-m^{2}=-2p_ik_f=-2m\omega_{f},
\end{equation}
\begin{equation}\label{DR2}
(p_i+k_i)^{2}-m^{2}=2p_ik_i=2m\omega_{i}.
\end{equation}

Secondly, we will find traces with the form $Tr[\slashed{A}_1\slashed{A}_2\ldots\slashed{A}_{n}]$ which are null in the case where $n$ is odd and in another one that we will present next. By using (\ref{TRAZ2}), it is not difficult to obtain the following result
\begin{equation}\label{T1}
Tr[\slashed{A}_1\slashed{A}_2\ldots\slashed{A}_{2n}]=(A_1.A_2)(A_3.A_4)\ldots(A_{2n-1}.A_{2n})+(A_2.A_3)(A_4.A_5)\ldots(A_{2n}.A_1).
\end{equation}

In consequence, from (\ref{pi epsilon}), (\ref{k epsilon}) and (\ref{T1}), we can construct many null traces. As an example we could write
\begin{equation}\label{TE2}
Tr[\ldots{\slashed{\varepsilon}_i\slashed{p}_i\slashed{\varepsilon}_i}\ldots]=0,
\end{equation}
\begin{equation}\label{TE3}
Tr[\ldots{\slashed{p}_i\slashed{\varepsilon}_f\slashed{k}_f}\ldots]=0,
\end{equation}
\begin{equation}\label{TE4}
Tr[\ldots{\slashed{\varepsilon}_f\slashed{p}_i\slashed{\varepsilon}_i}\ldots]=0,
\end{equation}
and other combinations.

Now, returning to (\ref{M}), we have
\begin{equation}\label{Mcompquadf}\begin{aligned}
|M_{a}|^2
&=\frac{e^4}{2^8m^4\pi^{4}\omega_{i}\omega_{f}p_i^0p_f^0}
\Bigg(Tr\bigg[\slashed{p}_f(\slashed{p}_f+m)\slashed{\varepsilon}_{f}\slashed{\varepsilon}_{i}\slashed{p}_i(\slashed{p}_i+m)\slashed{\varepsilon}_{i}\slashed{\varepsilon}_{f}\bigg]+\\
&\quad+Tr\bigg[\slashed{p}_f(\slashed{p}_f+m)\slashed{\varepsilon}_{f}\slashed{\varepsilon}_{i}\slashed{p}_i(\slashed{p}_i+m)\slashed{\varepsilon}_{f}\slashed{\varepsilon}_{i}\bigg]
+Tr\bigg[\slashed{p}_f(\slashed{p}_f+m)\slashed{\varepsilon}_{i}\slashed{\varepsilon}_{f}\slashed{p}_i(\slashed{p}_i+m)\slashed{\varepsilon}_{i}\slashed{\varepsilon}_{f}\bigg]+\\
&\quad+Tr\bigg[\slashed{p}_f(\slashed{p}_f+m)\slashed{\varepsilon}_{i}\slashed{\varepsilon}_{f}\slashed{p}_i(\slashed{p}_i+m)\slashed{\varepsilon}_{f}\slashed{\varepsilon}_{i}\bigg]\Bigg),\\
\end{aligned}\end{equation}
\begin{equation}\label{McompT}\begin{aligned}
\mathcal{M}^*_{a}\mathcal{M}_{b}
&=\frac{e^4}{2^8m^{5}\pi^{4}\omega_{i}\omega^{2}_{f}{p}^0_i{p}^0_f}Tr\bigg[(\slashed{\varepsilon}_{i}\slashed{\varepsilon}_{f}+\slashed{\varepsilon}_{f}\slashed{\varepsilon}_{i})\slashed{p}_f(\slashed{p}_f+m)\slashed{\varepsilon}_{i}(\slashed{p}_i-\slashed{k}_f)\times\\
&\quad\times(\slashed{p}_i-\slashed{k}_f+m)\boxed{\slashed{\varepsilon}_{f}}\slashed{p}_i(\slashed{p}_i+m)\bigg]
-\frac{e^4}{2^8m^{5}\pi^{4}\omega^{2}_{i}\omega_{f}{p}^0_i{p}^0_f}\times\\
&\quad\times{}Tr\bigg[(\slashed{\varepsilon}_{i}\slashed{\varepsilon}_{f}+\slashed{\varepsilon}_{f}\slashed{\varepsilon}_{i})
\slashed{p}_f(\slashed{p}_f+m)\slashed{\varepsilon}_{f}(\slashed{p}_i+\slashed{k}_i)(\slashed{p}_i+\slashed{k}_i+m)\boxed{\slashed{\varepsilon}_{i}}\slashed{p}_i(\slashed{p}_i+m)\bigg],\\
\end{aligned}\end{equation}
\begin{equation}\label{M2compconfeynf}\begin{aligned}
|\mathcal{M}_b|^2
&=\frac{e^4}{2^{8}m^4\pi^{4}\omega^{3}_{f}\omega_{i}}Tr\bigg[\slashed{p}_f(\slashed{p}_f+m)\slashed{\varepsilon}_{i}(\slashed{p}_i-\slashed{k}_f)\times\\
&\quad\times(\slashed{p}_i-\slashed{k}_f+m)\slashed{\varepsilon}_{f}\slashed{p}_i(\slashed{p}_i+m)\boxed{\slashed{\varepsilon}_{f}}(\slashed{p}_i-\slashed{k}_f)(\slashed{p}_i-\slashed{k}_f+m)\slashed{\varepsilon}_{i}\bigg]\\
&\quad-\frac{e^4}{2^{8}m^4\pi^{4}\omega^{2}_{f}\omega^{2}_{i}}Tr\bigg[\slashed{p}_f(\slashed{p}_f+m)\slashed{\varepsilon}_{i}(\slashed{p}_i-\slashed{k}_f)\times\\
&\quad\times(\slashed{p}_i-\slashed{k}_f+m)\boxed{\slashed{\varepsilon}_{f}}\slashed{p}_i(\slashed{p}_i+m)\slashed{\varepsilon}_{i}(\slashed{p}_i+\slashed{k}_i)(\slashed{p}_i+\slashed{k}_i+m)\slashed{\varepsilon}_{f}\bigg]\\
&\quad-\frac{e^4}{2^{8}m^4\pi^{4}\omega^{2}_{f}\omega^{2}_{i}}Tr\bigg[\slashed{p}_f(\slashed{p}_f+m)\slashed{\varepsilon}_{f}(\slashed{p}_i+\slashed{k}_i)\times\\
&\quad\times(\slashed{p}_i+\slashed{k}_i+m)\slashed{\varepsilon}_{i}\slashed{p}_i(\slashed{p}_i+m)\boxed{\slashed{\varepsilon}_{f}}(\slashed{p}_i-\slashed{k}_f)(\slashed{p}_i-\slashed{k}_f+m)\slashed{\varepsilon}_{i}\bigg]\\
&\quad+\frac{e^4}{2^{8}m^4\pi^{4}\omega_{f}\omega^{3}_{i}}Tr\bigg[\slashed{p}_f(\slashed{p}_f+m)\slashed{\varepsilon}_{f}(\slashed{p}_i+\slashed{k}_i)\times\\
&\quad\times(\slashed{p}_i+\slashed{k}_i+m)\slashed{\varepsilon}_{i}\slashed{p}_i(\slashed{p}_i+m)\boxed{\slashed{\varepsilon}_{i}}(\slashed{p}_i+\slashed{k}_i)(\slashed{p}_i+\slashed{k}_i+m)\slashed{\varepsilon}_{f}\bigg].\\
\end{aligned}\end{equation}

With the help of null traces combination that we constructed before, we can see that the boxed terms in (\ref{McompT}) and (\ref{M2compconfeynf}) are surrounded by other which cancel the terms that they belong, this is
\begin{equation}\label{nul}
\mathcal{M}^*_{a}\mathcal{M}_{b}=0=\mathcal{M}^{*}_{b}\mathcal{M}_{a}, \quad |\mathcal{M}_{b}|^{2}=0.
\end{equation}

Similarly, there are null terms in (\ref{Mcompquadf}) coming from the products $\slashed{\varepsilon}_i\slashed{p}_i\slashed{\varepsilon}_i$, $\slashed{\varepsilon}_f\slashed{p}_i\slashed{\varepsilon}_i$, $\slashed{\varepsilon}_i\slashed{p}_i\slashed{\varepsilon}_f$ and $\slashed{\varepsilon}_f\slashed{p}_i\slashed{\varepsilon}_f$. Avoiding this terms and those with an odd number of $\beta$-matrices, we have
\begin{equation}\label{Mcompquadf2}\begin{aligned}
|M_{a}|^2
&=\frac{e^4}{2^{8}m^4\pi^{4}\omega_{i}\omega_{f}p_i^0p_f^0}
\bigg(Tr\Big[\slashed{p}_f\slashed{p}_f\slashed{\varepsilon}_{f}\slashed{\varepsilon}_{i}\slashed{p}_i\slashed{p}_i\slashed{\varepsilon}_{i}\slashed{\varepsilon}_{f}\Big]+Tr\Big[\slashed{p}_f\slashed{p}_f\slashed{\varepsilon}_{f}\slashed{\varepsilon}_{i}\slashed{p}_i\slashed{p}_i\slashed{\varepsilon}_{f}\slashed{\varepsilon}_{i}\Big]\\
&\quad+Tr\Big[\slashed{p}_f\slashed{p}_f\slashed{\varepsilon}_{i}\slashed{\varepsilon}_{f}\slashed{p}_i\slashed{p}_i\slashed{\varepsilon}_{i}\slashed{\varepsilon}_{f}\Big]+Tr\Big[\slashed{p}_f\slashed{p}_f\slashed{\varepsilon}_{i}\slashed{\varepsilon}_{f}\slashed{p}_i\slashed{p}_i\slashed{\varepsilon}_{f}\slashed{\varepsilon}_{i}\Big]\bigg)\\
&=\frac{e^4}{2^{6}\pi^{4}\omega_{i}\omega_{f}mE_f}(\varepsilon_i.\varepsilon_f)^{2}\\
\end{aligned}\end{equation}

Replacing (\ref{Mcompquadf2}) and (\ref{nul}) into (\ref{seccion de choque diferencial comp}), we obtain
\begin{equation}\label{seccion de choque diferencial comp2}\begin{aligned}
\frac{d\sigma}{d\Omega}\Big|_{\text{lab}}
&=\frac{e^4\omega_f^2}{16\pi^{2}m^{2}\omega^{2}_i}(\varepsilon_i.\varepsilon_f)^{2}\\
&=\frac{\alpha^2\omega_f^2}{m^2\omega_i^2}\left(\varepsilon_{f}.{\varepsilon_{i}}\right)^2.\\
\end{aligned}\end{equation}

In the framework of Klein-Gordon-Fock equation, the result (\ref{seccion de choque diferencial comp2}) was obtained in \cite{Itz} and using CPT in \cite{Jhosep}. It is important to emphasize here that, as showed in (\ref{S(2)2}), the only non-null contribution for the differential cross section comes from the Dirac delta function which appear because of the order of singularity $\omega=0$ in the causal splitting process. In other words, as in the standard QFT approach, the only non-null contribution comes from the four legs vertex $\sim:\overline{\psi}(x)\slashed{A}_\mu(x)\slashed{A}_\nu(x)\psi(x):$ which appear from the theory in the CPT point of view.

\section{Discussion and Conclusions}\label{DC}

Along this article, we summarized and used the axiomatic approach for QFT known as CPT to compute the differential cross sections of scattering processes, in the tree level, for SDKP$_4$. The objective of this computation is compare the results with those obtained for SQED$_4$ to establish the equivalence between both formalisms.

In section \ref{PGI} we introduced the principle of \textit{perturbative gauge invariance} to determine the correct form of the base term $T_1(x_1)$ to construct the $S$-Matrix for SDKP$_4$. With the term $T_1$ defined, we determined the differential cross section in the scattering of a scalar particle via non quantized electromagnetic field obtaining the same result as that obtained in SQED$_4$. After that, we proceed to use CPT to determine de Causal 2-point distribution $D_{2}(x_1,x_2)$ which contain many processes: Moller, Bhabha, Compton, vacuum polarization and self-energy.

The differential cross section computed for Moller and Compton processes are the same for the ones obtained via SQED$_4$. We must highlight the case of Compton scattering where we found a singular DKP propagator $\omega=0$ because of the extra $\slashed{p}$ comming from the $\beta$-matrix algebra. The same happens in SQED$_4$ via CPT \cite{Scharf28,Jhosep} in the case where two derivatives scalar fields were contracted via Wick theorem. Furthermore, these singular propagators reproduce the second order terms $\sim{}e^{2}A_{\mu}A^{\mu}\phi^{*}\phi$ and $\sim{}e^{2}\bar{\psi}\slashed{A}_{\mu}\slashed{A}^{\mu}\psi$ for SQED$_4$ and SDKP$_4$, respectively. In SQED$_4$, via CPT or Feynman diagrams, the term $\sim{}e^{2}A_{\mu}A^{\mu}\phi^{*}\phi$ is really important because the contributions from the other diagrams are null \cite{Itz,Jhosep}.  For SDKP$_4$ we see the same in section \ref{S4}.

The recovery of the interaction between two scalars and two photons in the same point is the most relevant result of this work. We are completely sure that the interaction of four scalars, needed to have a renormalized theory, could be recovery too. As demonstrated in \cite{Scharf28}, for SQED$_4$, the latter comes naturally from the singular order $\omega>0$ of a general graph with four scalar legs. We let the latter computation for a future work. At least, in the tree level, we conclude that SDKP$_4$ and SQED$_4$ are equivalent.

\section*{acknowledgements}
J. Beltran thanks to CAPES for full support and B. M. Pimentel thanks to CNPq for partial support



\end{document}